\documentclass[aps,prb,twocolumn]{revtex4-1}

\usepackage{amsmath} 
\usepackage{amsthm} 
\usepackage{graphicx} 
\usepackage{float}
\usepackage{amssymb}
\usepackage{enumitem}
\begin{document}

\title{Tuning vortex fluctuations and the resistive transition in superconducting films with a thin overlayer }

\author{Alex Gurevich}
\email[]{gurevich@odu.edu}
\affiliation{Department of Physics, Old Dominion University, Norfolk, Virginia 23529, USA.}

\date{\today}

\begin{abstract}
It is shown that the temperature of the resistive transition $T_r$ of a superconducting film can be increased by a thin superconducting or normal overlayer. For instance, deposition of a highly conductive thin overlayer onto a dirty superconducting film can give rise to an "anti-proximity effect"  which manifests itself in an initial increase of $T_r(d_2)$ with the overlayer thickness $d_2$ followed by a decrease of $T_r(d_2)$ at larger $d_2$. Such a nonmonotonic thickness dependence of $T_r(d_2)$ results from the interplay of the increase of a net superfluid density mitigating phase fluctuations and the suppression of the critical temperature $T_c$ due to the conventional proximity effect. This behavior of $T_r(d_2)$ is obtained by solving the Usadel equations to calculate the temperature of the Berezinskii-Kosterletz-Thouless transition, and the temperature of the resistive transition due to thermally-activated hopping of single vortices in dirty bilayers. The theory incorporates relevant materials parameters such as thicknesses and conductivities of the layers, interface contact resistance between them and the subgap quasiparticle states which affect both phase fluctuations and the proximity effect suppression of $T_c$. The transition temperature $T_r$ can be optimized by tuning the overlayer parameters, which can significantly weaken vortex fluctuations and nearly restore the mean-field critical temperature.  The calculated  behavior of $T_r(d_2)$ may explain the nonmonotonic dependence of $T_r(d_2)$ observed on (Ag, Au, Mg, Zn)-coated Bi films, Ag-coated Ga and Pb films or NbN and NbTiN films on AlN buffer layers. These results suggest that bilayers can be used as model systems for systematic investigations of optimization of fluctuations in superconductors.  

\end{abstract}

\maketitle

%%%%%%%%%%%%%%%%%%
\section{Introduction}\label{section_introduction}

Recent discoveries of two-dimensional (2D) materials and interfaces with unique physical properties \cite{rev1,rev2,rev3,rev4,rev5,rev6},     
particularly, the observations of superconductivity in FeSe monolayers on strontium titanade \cite{fese1,fese2,fese3,fese4,fese5,fese6,fese7}, monolayers of Pb on Si substrates \cite{pb1,pb2,pb3} 
or 2H TaS$_2$ \cite{tas2} have 
renewed the interest in the pairing mechanisms and the effect of vortex fluctuations in extreme 2D superconductors.  In addition to the complex physics of charge transfer, strain effects and collective excitations at the interfaces, the observation of superconducting transition and the opening of the quasiparticle gap in FeSe monolayers at temperatures over 100 K brings about the following issue. The observed temperature of the resistive transition $T_r$ in a superconducting monolayer is always reduced by pairbreaking fluctuations of the order parameter and the Berezinskii-Kosterlitz-Thouless  (BKT) proliferation of vortices \cite{kt1,kt2}, which should be particularly  pronounced in dirty thin films like amorphous Pb monolayers \cite{pb1,pb2,pb3} or  FeSe monolayers with low superfluid density and the Fermi energies $E_F \simeq 10-100$ meV~ \cite{rev1,rev2,rev3,rev4,rev5,rev6}.  In that case a mean-field pairing temperature $T_c$ would be expected to be well above the observed $T_r\simeq 50$ K. The question is then what is the actual $T_c$ and to what extent it could restored by reducing fluctuations by materials nanostructuring.   

Pairbreaking fluctuations can be mitigated by enhancing the phase stiffness, which implies increasing the superfluid density or reducing the quasipaticle mass or electronic anisotropy \cite{pfluct,agrev}. It has been proposed to do so by combining strongly fluctuating superconducting layers with a nonsuperconducting materials with high carrier density \cite{comp1,comp2}. Using the Hubbard model for a superconducting (S) layer coupled to a normal (N) layer, it was shown that this mechanism can increase the phase stiffness in the bilayer and increase the transition temperature \cite{comp1,comp2}.  Yet testing this proposal experimentally would require a theory in which the observed $T_r$ in a bilayer is expressed in terms of accessible materials parameters such as thicknesses and conductivities of the S and N layers, and an interface contact resistance which can be readily tuned to optimize both the phase fluctuations and the proximity effect suppression of $T_c$. Such approach is developed in this work in which the resistive transition is associated with the BKT transition temperature $T_b$ or the temperature of the resistive transition caused by thermally-activated hopping of vortices. These transition temperatures were calculated here using the theory of proximity effect in dirty thin film bilayers  described by the Usadel equations \cite{cooper,degennes,golub,belzig,fominov,gol}. The theory shows that $T_r(d_2)$ first increases with the thickness of a conductive overlayer $d_2$, reaches a maximum which can be rather close to $T_c$ and then decreases as $d_2$ further increases. Such behavior of $T_r(d_2)$ resulting from the interplay of an enhanced phase stiffness and a reduction of $T_c$ due to the proximity effect, occurs if the conductivity of the overlayer is much higher than the conductivity of the S film in the normal state. In this case $T_r$ reaches maximum at the overlayer thicknesses much smaller than the thickness of the S film. 

The above mechanism may be relevant to the nonmonotonic dependencies of the resistive transition temperatures of ultra thin films on the thickness of conductive overlayers observed on (Ag, Au, Mg, Zn)-coated Bi films \cite{bi1,bi2}, Ag-coated Ga \cite{ga} and Pb films \cite{pb} or NbN and NbTiN films on AlN buffer layers \cite{nbn}.  It was also observed that $T_r$ of La$_{2-x}$Sr$_x$CuO$_4$ thin films capped by an overdoped metallic La$_{1.65}$Sr$_{0.35}$CuO$_4$ layer is higher than $T_c$ of the bare film, indicating the effect of enhanced phase stiffness \cite{koren}. 
Other experiments revealed the effect of disconnected metallic gates on $T_r$ of the 2D arrays of Al Josephson junctions \cite{jjatunee} and amorphous MoGe films \cite{kap}. Subsequent theories associated the effect of remote N overlayers on $T_r$ with a tunable dissipative environment affecting fluctuations of the order parameter which drive a superconductor-insulator transition \cite{jjatunet} and quantum tunneling of vortices \cite{finkels} though either capacitive or inductive coupling with the metallic gates.  It was also proposed to tune the BKT transition temperature with a decoupled thick S overlayer \cite{magtune}. Other mechanisms of the  nonmonotonic dependence of $T_r(d_2)$ may be related to a broader issue of interface superconductivity \cite{is1,is2} or the reduction of the Coulomb repulsion in the S film by a  thin N overlayer \cite{coulomb1,coulomb2}.  

In this work the effect of a thin overlayer on vortex fluctuations in a thin film is addressed, assuming that the overlayer is in contact with the film.  Here the effect of the overlayer on $T_r$ is associated with an increased energy of a perpendicular vortex. In this case restoring the mean-field $T_{c}$ could be achieved by depositing not only a highly conductive N overlayer but also a S overlayer with higher $T_c$ coupled through a Josephson buffer junction, for instance, a Bi-$2223$ or YBCO overlayer onto the FeSe monolayer. Such high-$T_c$ overlayer would be particularly effective to suppress the BKT fluctuations in a lower-$T_c$ layer.  Overlayers can also be used to reduce the effect of vortex fluctuations in granular films of arrays of Josephson junctions. A model developed here incorporates materials features into a theory of the BKT transition in a proximity coupled bilayer. This model primarilly focuses on the interplay of the phase stiffness and the proximity effect in the framework of a transparent single-vortex picture of the BKT transition, leaving aside a possibility of interface superconductivity and the effect of multi-vortex correlations on $T_b$. 

The paper is organized as follows. In Sec. II, the BKT transition in a dirty film is discussed, taking into account the effect of subgap states on $T_b$. In Sec. III restoration of the mean-field $T_{c}$ in solid and granular films covered with a high-$T_{c}$ overlayer is considered.  Sec. IV is devoted to the calculation of $T_{c}$ of S-N bilayers, taking into account the contact resistance and subgap states. In Sec. V reduction of the Ginzburg number and the effect of fluctuations on the transition temperature in a bilayer is addressed. In Sec. VI  a nonmonotonic dependence of the BKT transition temperature $T_b(d_2)$ on the thickness of a conductive N overlayer is calculated.  In Sec. VII finite size effects in the resistive transition caused by 
thermally-activated hopping of complete and fractional vortices in bilayers are considered. In Sec. VIII broader implications of the obtained results for the reduction of fluctuations in 2D superconductors are discussed.       

\section{BKT transition in a thin film}

This section gives a brief overview of the BKT transition temperature $T_b$ in dirty s-wave superconducting films for which the reduction of $T_b$ 
relative to the mean field critical temperature $T_{c}$ is most pronounced. Hereafter thin films 
with the Pearl magnetic penetration depth $\Lambda=\lambda_L^2/d_1\,$ \cite{pearl} larger than a lateral film size $L$ are considered, where $d_1$ is the film thickness, 
and $\lambda_L$ is the bulk London penetration depth.
   
\subsection{Non-granular films}

The BKT temperature is determined by the energy of a perpendicular vortex $\epsilon=\epsilon_0\ln(L/\xi)$ in a thin film \cite{kt1,kt2}:
\begin{equation}
\zeta\epsilon_0(T_b)=2T_b.
\label{tb}
\end{equation} 
Here the factor $\zeta < 1$ takes into account renormalization of the mean-field superfluid density by fluctuations (hereafter $T$ is measured in energy units). For instance, Monte Carlo simulations of vortices in the $XY$ model \cite{ren1,ren2,renexp} gave $\zeta=0.58$. In addition, $\zeta$ is reduced by weak localization effects in disordered films and amplitude fluctuations of the order parameter \cite{bktm,amplf}. 

The energy of the vortex $\epsilon$ in a thin film mostly comes from the kinetic energy of circulating currents. In the dirty limit $\epsilon$ is given by \cite{kopnin}
\begin{gather}
\epsilon = \int K(\textbf{r})d^2 \textbf{r}+\epsilon_c,
\label{eppo}\\
K(\textbf{r})=\frac{\pi\hbar\sigma_1 d_1 T}{2e^2}Q^2(\textbf{r})\sum_{\omega>0}\frac{\Delta^2}{\omega^2+\Delta^2},
\label{kino} 
\end{gather}
where $\mathbf{Q}=\nabla\chi+2\pi\mathbf{A}/\phi_{0}$ is proportional to the superfluid velocity, $\chi$ is the phase of the order parameter, $\mathbf{A}$ 
is the vector potential, $\sigma_1$ is a normal state conductivity, $\phi_0$ is the flux quantum, $e$ is the electron charge, $\epsilon_c\simeq 0.5\epsilon_0$ is a vortex core energy \cite{hu}, and $\Delta$ is the superconducting gap. 
Summing up over the Matsubara frequencies $\omega=\pi T(2n+1)$ and integrating in Eq. (\ref{eppo}) with $Q=1/r$ for a film with $\Lambda>L$ gives $\epsilon=\epsilon_0\ln(L/\xi)+\epsilon_c$, where
\begin{equation}
\epsilon_0=\frac{\pi\Delta R_0}{8 R}\tanh\frac{\Delta}{2T},\qquad R_0=\frac{h}{e^2}.
\label{epo}
\end{equation}
Here $R=(d_1\sigma_1)^{-1}$ is the sheet film resistance in the normal state, and $R_0=25.8$ kohm. Equations (\ref{tb}) and (\ref{epo}) combined with the BCS gap equation for $\Delta(T)$ form the basis for the calculations of $T_b$ in dirty films \cite{beasley}.  

This conventional approach does not take into account the essential effects of weak localization \cite{bktm}, inhomogeneities \cite{disordbkt,broad} and grain boundaries in polycrystalline films on $T_b$. Another relevant materials feature is the broadening of the gap singularities in the BCS density of states $N(\epsilon)$. Numerous STM experiments have shown that the DOS broadening can be significant, particularly in thin films and bilayers \cite{pb,tun,blstm1,blstm2,stmPb}. This effect is usually taken into account in the Dynes model \cite{dynes1,dynes2}:
\begin{equation}
N(\epsilon)=\mbox{Re}\frac{N_1(\epsilon+i\Gamma)}{\sqrt{(\epsilon+i\Gamma)^2-\Delta^2}},\quad \epsilon > 0.
\label{Nd}
\end{equation}  
Here $\Gamma$ quantifies a finite lifetime of quasiparticles $\sim \hbar/\Gamma$ resulting in subgap states at $\epsilon<\Delta$, and $N_1$ is the density of states in the normal state. Many mechanisms of subgap states have been considered in the literature, including inelastic scattering of quasiparticles on phonons \cite{inelast,kopnin}, Coulomb correlations \cite{coulomb}, anisotropy of the Fermi surface \cite{anis}, inhomogeneities of the BCS pairing constant \cite{larkin}, magnetic impurities \cite{balatski}, spatial correlations in impurity scattering \cite{balatski,meyer}, or diffusive surface scattering \cite{arnold}. 

The phenomenological Eq. (\ref{Nd}) captures the broadening of the DOS peaks at $\epsilon\approx \Delta$, but does not correctly describe low-energy tails in $N(\epsilon)$ obtained in microscopic calculations (see, e.g., Ref. \onlinecite{feigel} for an overview of different mechanisms). Details of exponential or power-law energy tails in $N(\epsilon)$ at $|\epsilon|\ll \Delta$ can be essential for the calculations of residual quasiparticle conductivity and surface resistance \cite{kubo}. However, vortex effects considered here are determined by the superfluid density which is weakly affected by the low-energy tails of $N(\epsilon)$ at $\epsilon\ll \Delta$. Thus, the conventional Eq. (\ref{Nd}) in which all microscopic mechanisms are included in a single parameter $\Gamma$ is rather useful to address the effect of the DOS broadening on the BKT transition by the simple substitution $\omega\to\omega+\Gamma$ in Eqs. (\ref{eppo})-(\ref{kino}). In this approach $\Gamma$ is regarded as a material parameter which can be extracted from tunneling measurements.  Then Eqs. (\ref{kino})-(\ref{epo}) yield
\begin{equation}
\epsilon_0=\frac{\Delta R_0}{4R}\mbox{Im}\psi \left[\frac{1}{2}+\frac{\Gamma}{2\pi T}+\frac{i\Delta}{2\pi T}\right],
\label{eg}
\end{equation} 
where $\psi(z)$ is a digamma function.  At $\Gamma=0$ Eq. (\ref{eg}) reduces to Eq. (\ref{epo}) since $\mbox{Im}\psi(1/2+ix)=(\pi /2)\tanh(\pi x)$. 
The equation for the pair potential $\Delta$ is given by 
\begin{equation}
\ln\frac{T_{c}}{T}=\sum_{n=0}^{\infty}\biggl[\frac{1}{n_1+\gamma}-
\frac{1}{\sqrt{(n_1+\gamma)^{2}+(\Delta/2\pi T)^2}}\biggr],
\label{gap0} 
\end{equation}
where $n_1=n+1/2$ and $\gamma=\Gamma/2\pi T$. The critical temperature is determined by the equation similar to 
that describes the reduction of $T_c$ by magnetic impurities \cite{balatski}:
\begin{gather}
\ln\frac{T_{c1}}{T_c}=U\left(\frac{\Gamma}{2\pi T_c}\right),
\label{tc1} \\
U(x)=\psi\left(\frac{1}{2}+x  \right)-\psi\left(\frac{1}{2}\right),
\label{U}
\end{gather}
where $T_{c1}=(2\gamma_E\Omega_1/\pi)\exp(-1/\lambda_1)$, $\lambda_1$ is a BCS pairing constant, $\Omega_1$ is the Debye frequency, and $\gamma_E=1.78$. 
Here $T_c$ vanishes at $\Gamma>\pi T_{c1}/\gamma_E$ and decreases linearly with $\Gamma$ at $\Gamma\ll 2\pi T_{c1}$:
\begin{equation}
T_c = T_{c1} - \frac{\pi\Gamma}{4}.
\label{tc2}
\end{equation} 
This equation may describe the reduction of $T_c$ in thin films due to the DOS broadening as the film thickness decreases, consistent with tunneling measurements \cite{pb,blstm1,blstm2,stmPb}.

\begin{figure}[tb]
   \includegraphics[width=9cm]{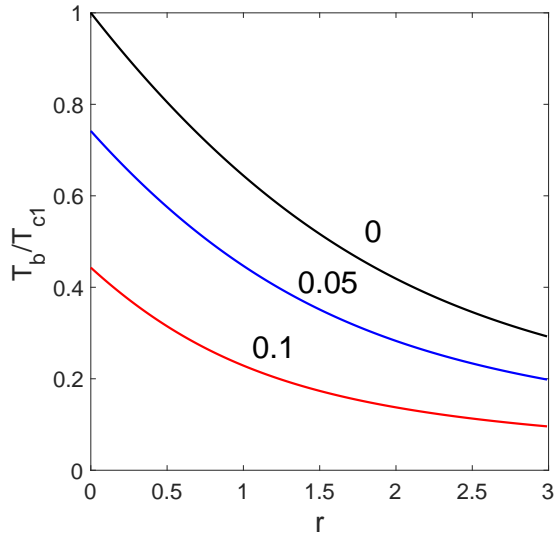}.
   \caption{ The BKT transition temperature as a function of the resistance ratio $r=8R/\pi\zeta R_0$ at different values of the DOS broadening parameter $\Gamma/2\pi T_{c1}$ calculated from 
   Eq. (\ref{bkt0}).
   }\label{fig1}
%   \vspace{-3mm}
\end{figure}

Combining Eqs. (\ref{tb}) and (\ref{eg}) yields the following equation for the BKT temperature $T_b(R)$: 
\begin{equation}
\frac{R}{R_0}=\frac{\zeta\Delta}{8T_b}\mbox{Im}\psi\left[\frac{1}{2}+\frac{\Gamma}{2\pi T_b}+\frac{i\Delta}{2\pi T_b}\right].
\label{bkt0}
\end{equation}
Shown in Fig. \ref{fig1} is $T_b(R)$ calculated from Eqs. (\ref{gap0}), (\ref{tc1}) and (\ref{bkt0}) 
for different values of the DOS broadening parameter $\gamma_1=\Gamma/2\pi T_{c1}$. Here the DOS broadening reduces the magnitudes of  
$T_c$ and $T_b$ but the overall dependence of the normalized $T_b/T_c$ on the sheet resistance does not change qualitatively as $\Gamma$ increases.     

\subsection{Granular films and Josephson junction arrays}

Granular films and Josephson junction arrays can be modeled by the energy functional of XY model \cite{jja1,jja2}
\begin{equation}
F=E_J\sum_{i\neq j}[1-\cos(\chi_i-\chi_j)],
\label{xy}
\end{equation}  
where the coupling energy $E_J=\hbar I_c/2e$ is proportional to the intergrain Josephson critical current $I_c$, and $\chi_j$ is the phase in the $j-$th grain. 
The energy of a vortex is then $\epsilon=\pi E_J\ln(L/a)$, where $a$ is a grain size.  For  SIS junctions,  
$I_c=(\pi\Delta/2eR_i)\tanh(\Delta/2T)$ is inversely proportional to the tunneling contact resistance $R_i$ between the grains \cite{KKL}, so that 
$\pi E_J=(\pi\Delta R_0/8 R_i)\tanh(\Delta/2T)$. For identical grain contacts, the equation for the BKT temperature $2T_b=\zeta\pi E_J$ thus becomes
\begin{equation}
\frac{R_i}{R_0}=\frac{\pi\zeta\Delta}{16T_b}\tanh\frac{\Delta}{2T_b},
\label{tbj}
\end{equation}
where $\Delta(T_b)$ is determined by Eqs. (\ref{gap0})-(\ref{tc1}), and the factor $\zeta<1$ takes into account mechanisms which reduce $I_c$ as compared to the BCS model, 
including fluctuations \cite{jja2} and materials factors which can result in $\zeta\simeq 0.2-0.8$ \cite{KKL}.  Equation (\ref{tbj}) coincides with Eq. (\ref{bkt0}) at $\Gamma=0$ for a non-granular film 
with the replacement $R_i\to R$.

For large $R_i$, the film sheet resistance $R=\alpha_i\bar{R}_i$ is proportional to a mean value $\bar{R}_i$, where the geometric constant $\alpha_i$ depends on  
spatial distribution of intergrain contacts, grain shapes, and distribution functions of intergrain areas $A_i$ and critical current densities \cite{granul,granrev}. The relations $R\propto \bar{R}_i$ 
and $\epsilon\propto R^{-1}$ no longer hold if the intergrain contacts are SNS Josephson junctions for which the $I_cR_i$ product can be much smaller than for SIS junctions \cite{degennes,golub}. Here  
the energy of the vortex $\epsilon_J=\pi\hbar I_c/2e$ and the BKT temperature 
can be greatly reduced by weakly-coupled SNS grain boundaries which do not necessarily result in high sheet 
resistance. 

\section{Weakly coupled overlayer}

\begin{figure}[tb]
	\includegraphics[width=10cm, trim={50mm 55mm 20mm 50mm},clip]{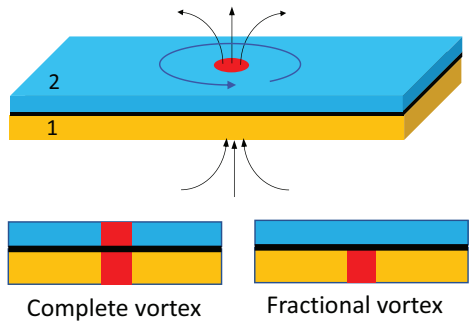}
   	\caption{ A perpendicular vortex in a superconducting bilayer. The horizontal black line represents either a weakly-coupled planar 
	Josephson junction or an interface with a sheet contact resistance $R_B$.  Bottom panel shows a complete core of a single-quantized vortex in a phase-locked bilayer 
	(left) and a partial core of a fractional vortex (right).  
   }\label{fig2}
%   \vspace{-3mm}
\end{figure}

Consider two superconducting layers separated by 
a planar Josephson junction with the critical current density $J_c$, as shown in Fig. \ref{fig2}. Let the layers 1 and 2 have the critical temperature 
$T_{c1}$ and $T_{c2}>T_{c1}$, and the gaps $\Delta_1$ and $\Delta_2$ be unaffected by weak Josephson coupling.  
The energy of a perpendicular vortex depends crucially on whether both layers are in a phase-locked state with $\chi_1(\textbf{r})=\chi_2(\textbf{r})$ or 
in a phase-unlocked state with different phases of the order parameter $\chi_1(\textbf{r})$ and $\chi_2(\textbf{r})$ in the layer 1 and 2. In the first case the vortex core threads both layers which thus have the same 
distribution of $Q(r)$. In a phase-unlocked bilayer a fractional vortex with 
a partial vortex core which threads only a lower-$T_c$ layer 1 can occur. The fractional vortex has a smaller kinetic energy of supercurrents in the layer 2 but
it produces the interlayer phase difference, $\chi =\chi_2-\chi_1$ and thus the Josephson energy $W_J = (\hbar J_c/2e)\int (1-\cos\chi)dxdy \sim \hbar J_c Lw/2e$ proportional to the area of the bilayer of length $L$ and width $w$, as shown in Appendix \ref{ap3}. For instance, if $T_{c1}$ and $T_{c2}$ are not very different, 
$J_c=\pi\Delta_1\Delta_2/4eR_\perp T_{c1}$ at $T\approx T_{c1}$, where $R_\perp$ is the interface resistance per unit area \cite{golub}.  The energy difference $\Delta W$ between the partial and the 
complete vortex is then:
\begin{equation}
\Delta W\simeq \frac{\Delta_1\Delta_2R_0}{16 T_{c1}R_\perp} Lw-\frac{\pi\Delta_2^2 R_0}{16T_{c2}R_2}\ln\frac{w}{\xi_2},
\label{dwf}
\end{equation}   
where $R_2=(d_2\sigma_2)^{-1}$ is the sheet resistance of layer 2.  The first term in Eq. (\ref{dwf}) describes the loss of the Josephson energy 
in a phase unlocked bilayer, and the second term is the gain in the kinetic energy in the layer 2. 
The complete vortex is more energetically favorable in wide films or long bridges in which $\Delta W>0$ and 
\begin{equation}
L > L_c \simeq (\pi R_\perp \Delta_2/w\Delta_1R_2)\ln(w/\xi_2).
\label{lcc}
\end{equation}
Fractional vortices may occur in narrow short bridges with $L<L_c$, particularly at $T\to T_{c1}$ where $\Delta_1(T)/\Delta_2\lesssim (R_\perp/R_2Lw)\ln(w/\xi)$. Here we focus on the BKT transition due to proliferation of complete vortices.

 The energy of a complete vortex is a sum of kinetic energies of currents in the layers 1 and 2 given by Eq. (\ref{epo}) for negligible DOS broadening. In this case  
 the equation for the BKT temperature takes the form:   
 \begin{equation}
 \frac{R_1}{R_0}=\frac{\pi\zeta}{16T_b}\left[\Delta_1\tanh\frac{\Delta_1}{2T_b}+ \frac{d_2\sigma_2}{d_1\sigma_1}\Delta_2\tanh\frac{\Delta_2}{2T_b}\right].
 \label{tbl}
 \end{equation}   
 As the overlayer thickness $d_2$ increases, $T_b$ increases and 
 exceeds $T_{c1}$ of the layer 1 if:
 \begin{equation}
 d_2>d_{2c}=\frac{16T_{c1}}{\pi\zeta\Delta_2(T_{c1})\sigma_2R_0}\coth\frac{\Delta_2(T_{c1})}{2T_{c1}},
 \label{alk}
 \end{equation} 
where $\Delta_1(T_b)=0$. As $d_2$ approaches $d_{2c}$, the fractional vortex becomes more energetically favorable.
Yet the high-$T_c$ overlayer restores the mean field $T_{c1}$ in the layer 1 
by increasing the sheet superfluid density and suppressing the BKT proliferation of vortices.

\section{Proximity-coupled overlayer}

In this section we follow the well-established theory of $T_c$ in a dirty thin film bilayer \cite{degennes,golub,belzig,fominov, gol} and take into account the effect of the DOS broadening essential in the subsequent analysis. 
A dirty bilayer comprising a superconductor $1$ at $-d_1<x<0$ and a superconductor $2$ at $0<x<d_2$ can be described by the Usadel equations:
\begin{eqnarray}
-D_1\theta_1^{''}+2\omega\sin\theta_1=2\varDelta_1\cos\theta_1,
\label{u1} \\
-D_2\theta_2^{''}+2\omega\sin\theta_2=2\varDelta_2\cos\theta_2,
\label{u2}
\end{eqnarray}
where  $D_1$ and $D_2$ are electron diffusivities in the layer 1 and 2, respectively, and   
\begin{equation}
\Delta_{1,2}=2\pi T\lambda_{1,2}\sum_{\omega>0}^{\Omega_{1,2}}\sin\theta_{1,2}.
\label{gap12}
\end{equation}
Here ($\lambda_{1}$, $\Omega_{1}$) and ($\lambda_2$, $\Omega_2$) are the pairing constant and the Debye frequency in a  
superconductor $1$ and $2$, respectively. 
Equations (\ref{u1}) and (\ref{u2}) are supplemented by the boundary conditions \cite{golub}:
\begin{gather}
\sigma_2\theta_2^{'}(0)=\sigma_1\theta_1^{'}(0)=R_B^{-1}\sin(\theta_1-\theta_2),
\label{bc1} \\
\theta'_1(-d_1)=\theta_2^{'}(d_2)=0,
\label{bc2}
\end{gather}
where $R_B$ is the sheet contact resistance of the interface. The DOS broadening is taken into account by $\omega\to \omega_1=\omega+\Gamma_1$ in Eq. (\ref{u1}) and $\omega\to\omega_2=\omega+\Gamma_2$ in Eq. (\ref{u2}). 

In the paper a thin film Cooper limit is considered, in which $d_{1,2}\ll (\hbar D_{1,2}/2\pi T_c)^{1/2}$ so that     
$\theta_1(x)$ and $\theta_2(x)$ are nearly constant across the layers \cite{cooper}. In this case 
the solution of Eqs. (\ref{u1}) and (\ref{u2}) given in Appendix \ref{ap1} yields two coupled equations for $\theta_1$ and $\theta_2$:
\begin{gather}
\tan\theta_2=\frac{\sin\theta_1+\alpha\beta\Delta_2}{\cos\theta_1+\alpha\beta\omega_2},
\label{t12} \\
\Delta_1\cos\theta_1-\omega_1\sin\theta_1=
\nonumber \\
\!\!\!\frac{\alpha(\omega_2\sin\theta_1-\Delta_2\cos\theta_1)}{\sqrt{1+\alpha^2\beta^2(\omega_2^2+\Delta_2^2)+2\alpha\beta(\omega_2\cos\theta_1+\Delta_2\sin\theta_1)}},
\label{tt1}
\\
\alpha =\frac{d_2N_2}{d_1N_1} , \qquad\quad \beta =4d_1N_1e^2R_B.
\label{am}
\end{gather}
General solutions of Eqs. (\ref{gap12}), (\ref{t12}) and (\ref{tt1}) can be obtained numerically. For   
a negligible contact resistance, $\alpha\beta\Omega_{1,2} \ll 1$, Eqs. (\ref{t12}) and (\ref{tt1}) yield  
$\theta_1=\theta_2\equiv \theta$, and the bilayer is described by the composite parameters:
\begin{gather}
\sin\theta=\frac{\Delta}{\sqrt{(\omega+\Gamma)^{2}+\Delta^{2}}},
\label{Sin} \\
\Delta=\frac{d_1N_1\Delta_1+d_2N_2\Delta_2}{d_1N_1+d_2N_2},
\label{del} \\
\Gamma=\frac{d_1N_1\Gamma_1+d_2N_2\Gamma_2}{d_1N_1+d_2N_2}.
\label{gam}
\end{gather}
The critical temperature $T_{c0}$ of the bilayer is obtained by linearizing Eqs. (\ref{u1})-(\ref{gap12}) with respect to $\theta_{1,2}\ll 1$:
\begin{equation}
T_{c0}=T_{c1}\exp\left[\frac{\alpha(\lambda_2-\lambda_1+\lambda_1\lambda_2\ln(\Omega_2/\Omega_1))}{(\lambda_1+\alpha\lambda_2)\lambda_1}\right],
\label{Tc}
\end{equation}
where $T_{c1}=(2\gamma_E\varOmega_1/\pi)\exp(-1/\lambda_1)$ is the critical temperature of the superconductor 1 with $\lambda_1>\lambda_2$.
The equation for $T_c$ in a bilayer with $\Gamma>0$ and $R_B=0$ reduces to Eq. (8) in which $T_{c1}\to T_{c0}$, and $\Gamma$ and $T_{c0}$ 
are given by Eqs. (28) and (29).  

\begin{figure}[tb]
   \includegraphics[width=8.5cm]{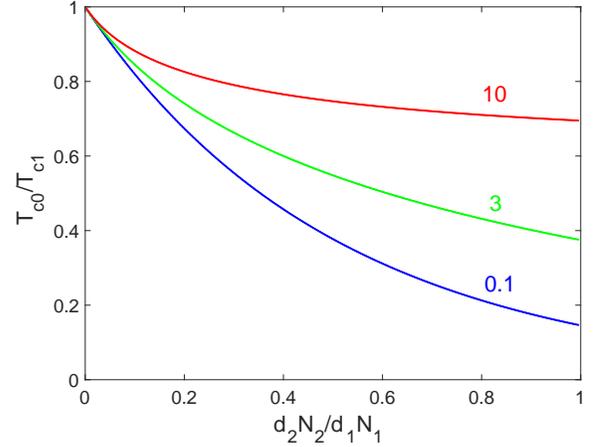}
   \caption{
Critical temperature $T_{c0}(d_2)$ of the N-S bilayer  
calculated from Eq. (\ref{Tcc}) for $\lambda_1=0.5$, $\lambda_2=0$, and different values of the contact resistance parameter $2\pi \beta T_{c1}=0.1, 3, 10$. 
   }\label{fig3}
     \vspace{-3mm}
\end{figure}

A general equation for $T_c$ at arbitrary $R_B$ was obtained in Appendix \ref{ap1}.
For a bilayer comprising a normal overlayer with $\lambda_2=0$, this equation simplifies to 
\begin{equation}
\frac{1}{\lambda_1}=2\pi T_c\sum_{\omega>0}\frac{(1+\alpha\beta\omega_2)}{[\omega_1(1+\alpha\beta\omega_2)+\alpha\omega_2]}\frac{\Omega_1^2}{(\omega^2+\Omega_1^2)}.
\label{tcc}
\end{equation}
Here the ad-hoc factor $\Omega_1^2/(\omega^2+\Omega_1^2)$ provides convergence of the sum for any relation between $\alpha\beta$ and 
$\Omega_1$, reproducing the BCS results while eliminating artifacts coming from the hard cutoffs in the sums at $\omega=\Omega_{1}$ in realistic cases of not very large $\Omega_{1}/2\pi T_c$. 
If $\Gamma_1$ and $\Gamma_2$ are negligible, Eq. (\ref{tcc}) becomes (see Appendix \ref{ap1}):
\begin{gather}
\ln\frac{T_{c}}{T_{c0}}=\frac{\alpha(\alpha\beta\Omega_1)^{2}}{(\alpha\beta\Omega_1)^{2}+(1+\alpha)^{2}}\bigg[\ln\frac{2\gamma_{E}\Omega_1}{\pi T_{c}}+
\nonumber \\
\frac{\pi(1+\alpha)}{2\alpha\beta\Omega_1}-U\left(\frac{1+\alpha}{2\pi\alpha\beta T_{c}}\right)\bigg],
\label{Tcc} 
\end{gather}   
where $T_{c0}=T_{c1}\exp(-\alpha/\lambda_1)$ is the critical temperature of the bilayer with $\beta\propto R_B=0$.
The contact resistance weakens the proximity effect coupling of the S and N layers, ameliorating the decrease of $T_c$ with $d_2$, as shown in Fig. \ref{fig3}.  The strongest proximity effect suppression 
of $T_c$ described by Eq. (\ref{Tc}) occurs at $R_B=0$. At nonzero contact resistance, $T_c(d_2)$ does not vanish at $d_2\to\infty$ but approaches a minimum value $T_{min}$ which increases with $R_B$  
so that $T_{min} \to T_{c1}$ at $\alpha\beta\Omega_1\gg 1$.

\begin{figure}[tb]
   \includegraphics[width=8.5cm]{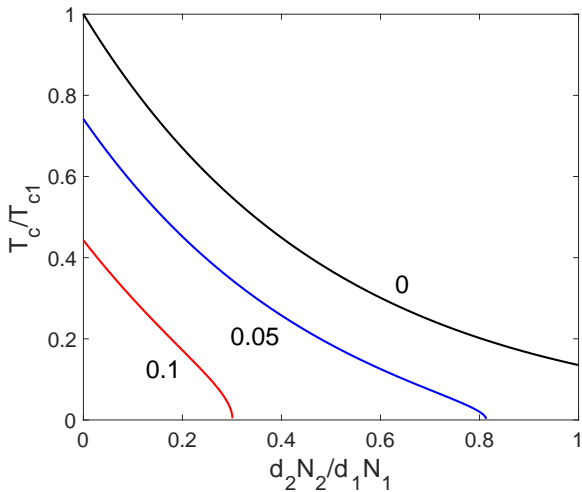}
   \caption{
Critical temperature $T_c(d_2)$ of the N-S bilayer  
calculated from Eqs. (\ref{tc1}), (\ref{U}), (\ref{gam}) and (\ref{Tc}) for $\lambda_1=0.5$, $\lambda_2=0$,  and different values of $\gamma_1=\Gamma/2\pi T_{c1}$. 
   }\label{fig4}
%   \vspace{-3mm}
\end{figure}

Figure \ref{fig4} shows the effect of DOS broadening on $T_c(d_2)$ in a N-S bilayer with $R_B=0$ and $\lambda_2=0$ calculated from Eqs. (\ref{tc1}) and (\ref{U}) for  
different values of $\gamma_1=\Gamma/2\pi T_{c1}$. Here the DOS broadening causes a stepper decrease of $T_c$ with $d_2$, the critical temperature vanishes if $d_2>d_c$. Setting $T_c\to 0$ and using $\psi(z)=\ln z$ at $z\gg 1$ in Eqs. (\ref{tc1}) and (\ref{U}) yields the following equation for $d_c$:
\begin{equation}
\Gamma(d_c)=\pi T_{c0}(d_c)/\gamma_E,
\label{critd}
\end{equation}
which has the same form as the equation for the critical concentration of paramagnetic impurities in the Abrikosov-Gorkov theory \cite{balatski}. 

\section{Fluctuations in N-S bilayers}

For a phase-locked N-S bilayer with $R_B=0$, the Ginzburg-Landau (GL) free energy functional is given by (see Appendix \ref{ap2}): 
\begin{gather}
\!\!\!F=\!\int \bigl[a(T)|\Psi|^2+c\bigl|\bigl(\nabla-\frac{2\pi i{\bf A}}{\phi_0}\bigr)\Psi\bigr|^2+\frac{b}{2}|\Psi |^4\bigr] d^2{\bf r},
\label{gl} \\
a=\frac{(T_{c0}-T)\nu}{T_{c0}}, \qquad b=\frac{7\zeta(3)\nu}{8\pi^2T_{c0}^2},
\label{ab} \\
c=\frac{\pi\hbar}{8T_{c0}}(d_1N_1D_1+d_2N_2D_2),
\label{c}
\end{gather}
where $\Psi=\Delta e^{i\chi}$ is the order parameter. The mean field jump in the specific heat $\Delta C=\nu^2/bT_{c0}$ at $T_{c0}$ is then:
\begin{equation}
\Delta C=\frac{8\pi^2\nu T_{c0}}{7\zeta(3)},\qquad \nu=d_1N_1+d_2N_2.
\label{dC}
\end{equation}
The Gaussian fluctuation correction to the sheet specific heat \cite{fluct} is readily obtained from Eq. (\ref{gl}):
\begin{equation}
\delta C(T)=\frac{2\nu e^2T_{c0}^2}{\pi^2\hbar(d_1\sigma_1+d_2\sigma_2)(T_{c0}-T)},\quad T>T_{c0}.
\label{flC}
\end{equation}
The width of the critical region 
of strong fluctuations $T_f-T_{c0}$, where $\delta C(T_f)=\Delta C$ defines the Ginzburg parameter $Gi=(T_f-T_{c0})/T_{c0}$ given by:
\begin{equation}
Gi=\frac{7\zeta(3)e^2}{4\pi^4 \hbar(d_1\sigma_1+d_2\sigma_2)}=\frac{7\zeta(3)R_\square}{2\pi^3R_0}.
\label{Gi}
\end{equation}
Here $Gi$, controlled by the ratio of the bilayer normal sheet resistance $R_\square=(d_1\sigma_1+d_2\sigma_2)^{-1}$ and the quantum 
resistance $R_0=h/e^2$, does not depend on superconducting properties \cite{fluct}.  A thin overlayer with $\sigma_2\gg\sigma_1$ and $d_2>d_1\sigma_1/\sigma_2\ll d_1$   
can thus strongly reduce $Gi$ and mitigate fluctuations without a significant suppression of $T_{c0}$ due to the proximity effect. 

The GL coherence length $\xi$ is defined here by the condition $a\Delta^2\simeq c\Delta^2/\xi^2$, giving
\begin{equation}
\xi=\left[\frac{\pi\hbar(d_1\sigma_1+d_2\sigma_2)}{16\nu e^2|T_{c0}-T|}\right]^{1/2}.
\label{xi}
\end{equation}
Generally, the global phase coherence is lost at a transition temperature $\tilde{T}_c$ at which the thermal energy $T$ is of the order of the condensation energy 
$\pi a^2\xi^2/2b$ within a correlated area $\pi\xi^2$, that is, $\mu_1\tilde{T}_c= a^2(\tilde{T_c})\pi\xi^2(\tilde{T}_c)/2b$, where $\mu_1\sim 1$. Using here Eqs. (\ref{ab}) and (\ref{xi}) yields:
\begin{equation}
\tilde{T}_c=\frac{T_{c0}(d_2)}{1+\mu R_\square(d_2)/R_0},
\label{tilTc}
\end{equation} 
where $\mu=56\zeta(3)\mu_1/\pi^3$. For instance, the BKT transition corresponds to $\mu_1\simeq 1/2$ and $\mu\simeq 1.1$. 
Fluctuations reduce $\tilde{T}_c$ relative to $T_{c0}$, but as the overlayer thickness increases, the effect of fluctuations weakens while $T_{c0}(d_2)$ gets diminished by the proximity effect. 
If $\sigma_2\gg\sigma_1$,  the transition temperature $\tilde{T}_c(d_2)$ first increases with $d_2$ due to decreasing $R_\square(d_2)$ in Eq. (\ref{tilTc}) and 
then decreases at larger $d_2$ as the proximity effect takes over. The nonmonotonic $\tilde{T}_c(d_2)$ occurs if $\partial\tilde{T}_c/\partial d_2>0$ at $d_2\to 0$, which  
in the case of $R_B=0$ and $\lambda_2=0$ reduces to:
\begin{equation}
\frac{D_2}{D_1}>\frac{1}{\lambda_1}\left(1+\frac{R_0}{\mu R}\right).
\label{cond}
\end{equation}
This inequality can be satisfied for a highly conductive N overlayer with $q=D_2/D_1\gg 1$. Here  
the maximum $\tilde{T}_c$ defined by Eqs. (\ref{Tc}) and (\ref{tilTc}) occurs at $\alpha_m=(\mu R/R_0\lambda_1 q)^{1/2}\ll 1$, and the optimum overlayer 
thickness $d_{2m}$ and the transition temperature $\tilde{T}_c(d_{2m})=T_{c1}(1-2\alpha_m/\lambda_1)$ become: 
\begin{gather}
d_{2m}=\frac{d_1N_1}{N_2}\left(\frac{\mu\lambda_1 R}{R_0 q} \right)^{1/2},
\label{d2m} \\
\tilde{T}_{c}(d_{2m}) =T_{c1}\left(1-2\sqrt{\frac{\mu R}{\lambda_1 q R_0}}\right).
\label{topt}
\end{gather}
At  $q=D_2/D_1\ll 1$ the optimum  overlayer thickness $d_{2m}$ is much smaller than the thickness of the S film,  
neither $d_{2m}$ nor $\tilde{T}_{c}(d_{2m})$ depending on $D_1$. 
Such N overlayer can nearly 
restore $\tilde{T_c}$ to the mean-field $T_{c1}$ of the S film.
Equations (\ref{gl})-(\ref{c}) do not take into account renormalization of   
the GL coefficients due to strong electron-phonon coupling \cite{lam1,lam2,lam3} and weak localization effects which become essential for large $R$~ \cite{bktm}. These effects influence the numerical factor $\mu$ but do not change the 
conclusion that a thin, highly conductive overlayer mitigates superconducting fluctuations.    

\section{BKT transition in a bilayer}

The interplay of the proximity effect and the phase stiffness manifests itself in the BKT transition temperature which shows 
how $T_r$ is affected by a thin overlayer. Here the vortex energy scale $\epsilon_0$ in Eq. (\ref{tb}) is determined by  
the sum of kinetic energies of circulating currents in the phase-locked layers 1 and 2:
\begin{equation}
\epsilon_0=\frac{\pi^2\hbar T}{e^2}\sum_{\omega>0} [d_1\sigma_1\sin^2\theta_1
+d_2\sigma_2\sin^2\theta_2]. 
\label{kin}
\end{equation}

Calculation of $T_b(\alpha)$ in the general case when $R_B$ is essential requires numerical solution of coupled Eqs. (\ref{tb}), (\ref{t12}), (\ref{tt1}) and (\ref{kin}).   
The behavior of $T_b(\alpha)$ becomes more transparent in a bilayer with a negligible $R_B$ for which the enhancement of the phase stiffness by the overlayer is most pronounced. In this case $\theta_1=\theta_2\equiv \theta$  is given by Eq. (\ref{Sin}), and Eq. (\ref{kin}) becomes
\begin{gather}
\epsilon_0=\frac{\pi R_0}{2R} (1+q\alpha)S, \qquad q=\frac{D_2}{D_1},
\label{epso} \\
S=T\sum_{\omega>0}\sin^{2}\theta=\frac{\Delta}{2\pi}\mbox{Im}\psi\left[\frac{1}{2}+\gamma+\frac{i\Delta}{2\pi T}\right].
\label{S}
\end{gather}
Here $\gamma=\Gamma/2\pi T$, and $S=(\Delta/4)\tanh(\Delta/2T)$ at $\gamma=0$. The vortex core radius $\simeq \xi$ given by Eq. (\ref{xi}) can be significantly increased by a  
highly conductive overlayer. 

Using Eqs. (\ref{tb}) and (\ref{epso}) the equation for the BKT temperature $T_b$ can be written in the form:
\begin{equation}
\frac{R}{R_0}=\frac{\zeta\Delta}{8T_b}(1+q\alpha)\mbox{Im}\psi\left[\frac{1}{2}+\frac{\Gamma}{2\pi T_b}+\frac{i\Delta}{2\pi T_b}\right],
\label{bkt}
\end{equation}
Here $T_b$ and the composite gap parameter $\Delta$ as functions of the film sheet resistance $R=(\sigma_1d_1)^{-1}$ and the overlayer thickness are determined self-consistently by Eqs (\ref{gap0}), (\ref{del}), (\ref{gam}), and (\ref{bkt}). The factor $\zeta$ accounts for the renormalization of the superfluid density and diffusivities due to strong electron-phonon coupling \cite{lam1,lam2,lam3}, 
fluctuations and weak localization effects \cite{bktm}. Given  
the complexity of the theoretical account of these mechanisms in bilayers affected by many uncertain microscopic parameters, 
$\zeta$ is treated here as a material parameter which can be expressed in terms of the observed $T_b$ in a single S film \cite{comm}.  

If $\Gamma=0$, the equations for $T_b$ can be written in the convenient parametric form:
\begin{gather}
q\alpha=\frac{r}{p}\coth p-1, \quad\qquad r=\frac{8R}{\pi\zeta R_0},
\label{y} \\
\ln\frac{T_{c1}}{T_b}-\frac{\alpha[\lambda_1-\lambda_2-\lambda_1\lambda_2\ln(\Omega_2/\Omega_1)]}{(\lambda_1+\alpha\lambda_2)\lambda_1}=
\nonumber \\
\sum_{n=0}^{\infty}\biggl[\frac{1}{n+\frac{1}{2}}-\frac{1}{\sqrt{(n+\frac{1}{2})^2+(p/\pi)^2}}\biggr].
\label{ktb} 
\end{gather}

\begin{figure}[tb]
   \includegraphics[width=8cm]{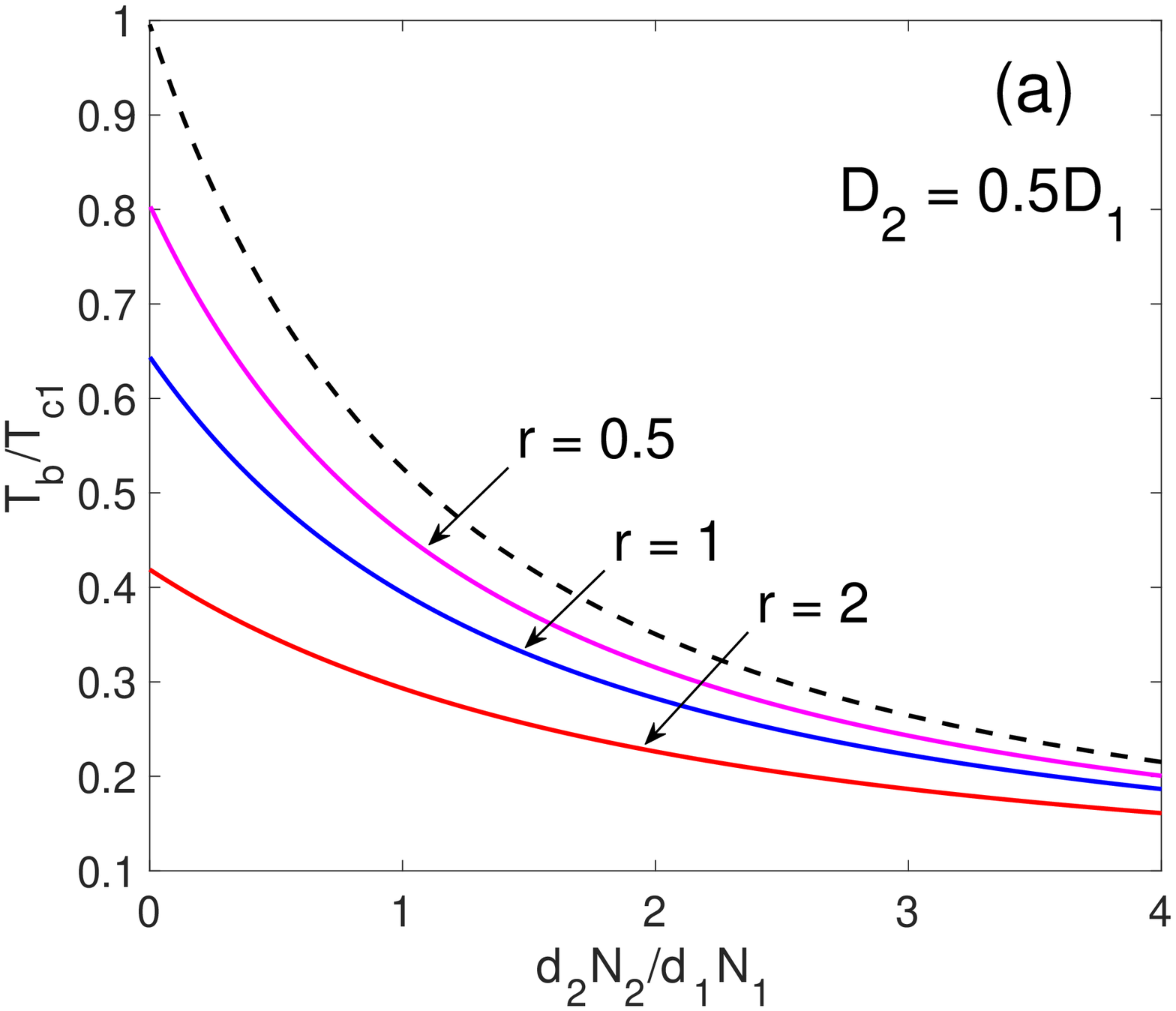}
    \\ 
   \includegraphics[width=8cm]{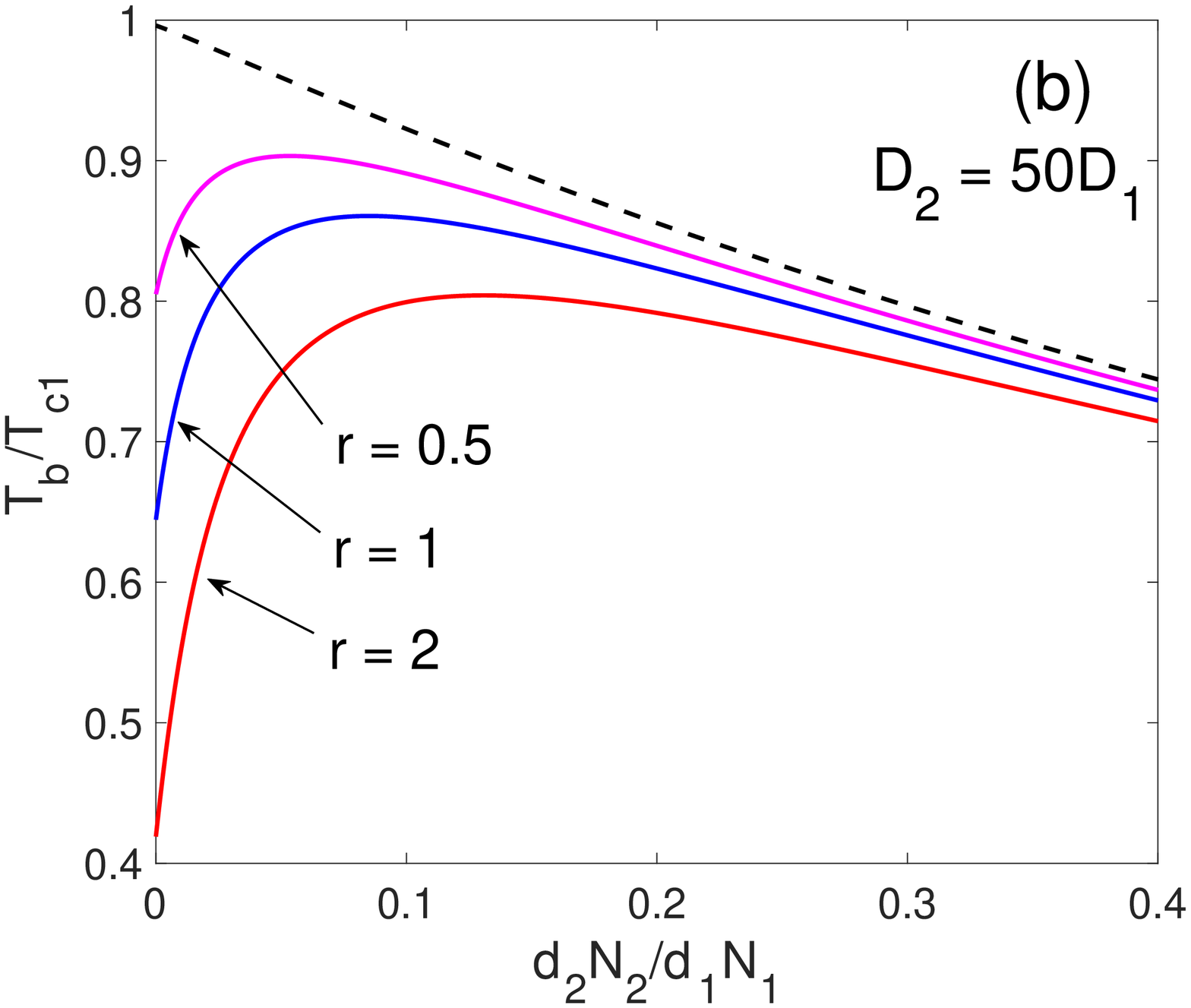}
   \caption{ BKT transition temperature $T_b(d_2)$ calculated from Eqs. (\ref{y})-(\ref{ktb}) for different film resistances, $r=8R/\pi\zeta R_0$, $\lambda_1=0.7$, $\lambda_2=0.2$, $\Omega_2=2\Omega_1$, and (a) $D_2=0.5D_1$ and (b) $D_2=50D_1$. The dashed line shows the proximity effect-limited $T_{c0}(d_2)$ in the absence of the BKT fluctuations. 
   }\label{fig5}
\end{figure}

Shown in Fig. \ref{fig5} is $T_b(\alpha)$ calculated from Eqs. (\ref{y}) and (\ref{ktb}) for different resistance parameters $r$  as the parameter $p=\Delta/2 T_b$ increases from $0$ to $\infty$. 
The behavior of $T_b(\alpha)$ depends essentially on the diffusivity ratio $q=D_2/D_1$. At $q\lesssim 1$, both $T_{c0}(\alpha)$ and $T_b(\alpha)$ decrease with the overlayer thickness 
in a way expected from the proximity effect, the difference between $T_b(\alpha)$ and $T_{c0}(\alpha)$ increasing with $r$. However, if $q\gg 1$,  the BKT temperature 
$T_b(\alpha)$ first increases with $d_2$ reaching a maximum at $d_2\ll d_1$ and then approaches $T_{c0}(\alpha)$ at larger $d_2$ as shown in Fig. \ref{fig5}b. This nonmonotonic $T_b(\alpha)$ at $q\gg 1$ results from the interplay of the increasing sheet superfluid density and the decreasing $T_{c0}$ due to the proximity effect, as was discussed in the previous section.    
\begin{figure}[tb]
   \includegraphics[width=11.5cm, trim={80mm 0mm 40mm 0mm},clip]{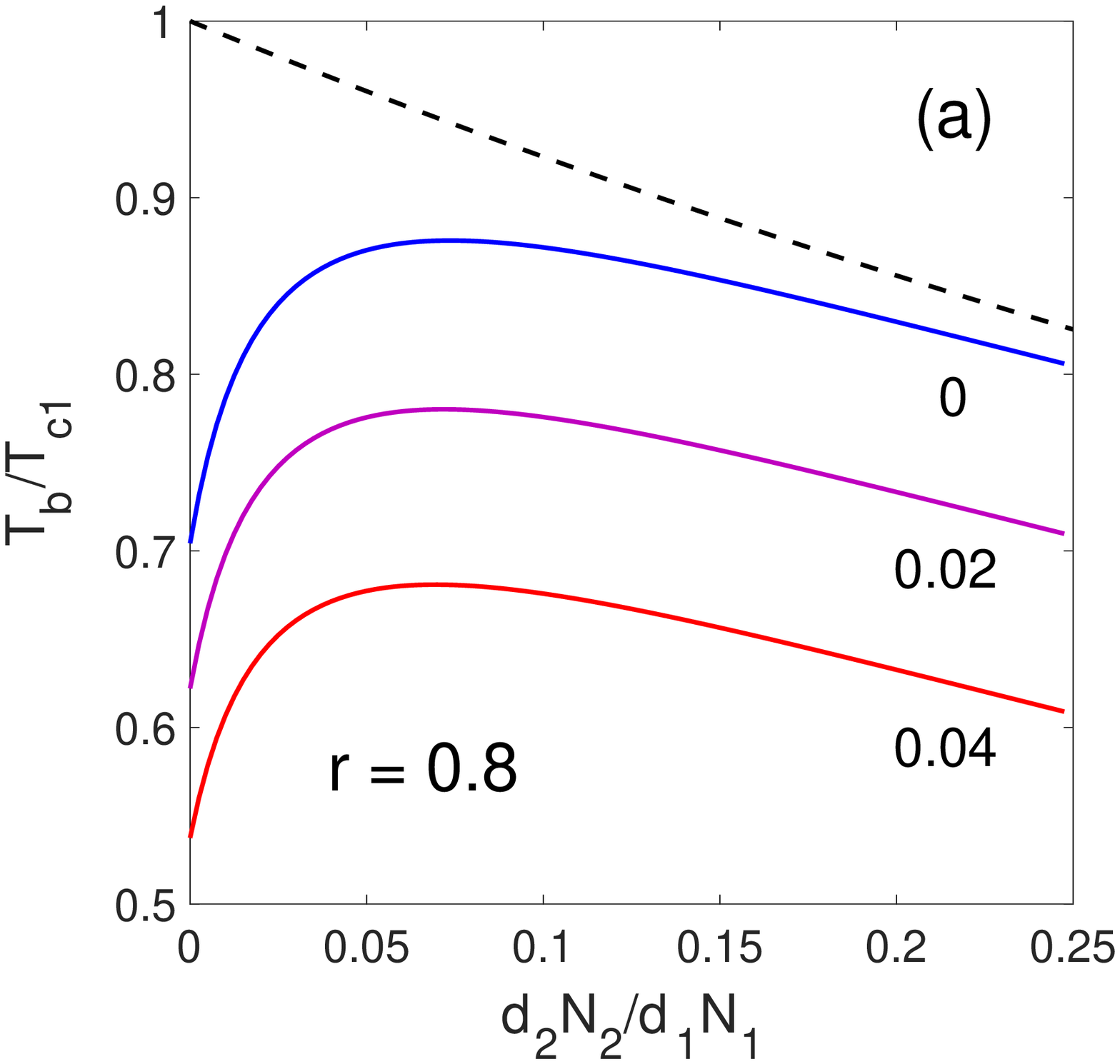}
    \\ 
    \includegraphics[width=11.5cm, trim={80mm 0mm 40mm 0mm},clip]{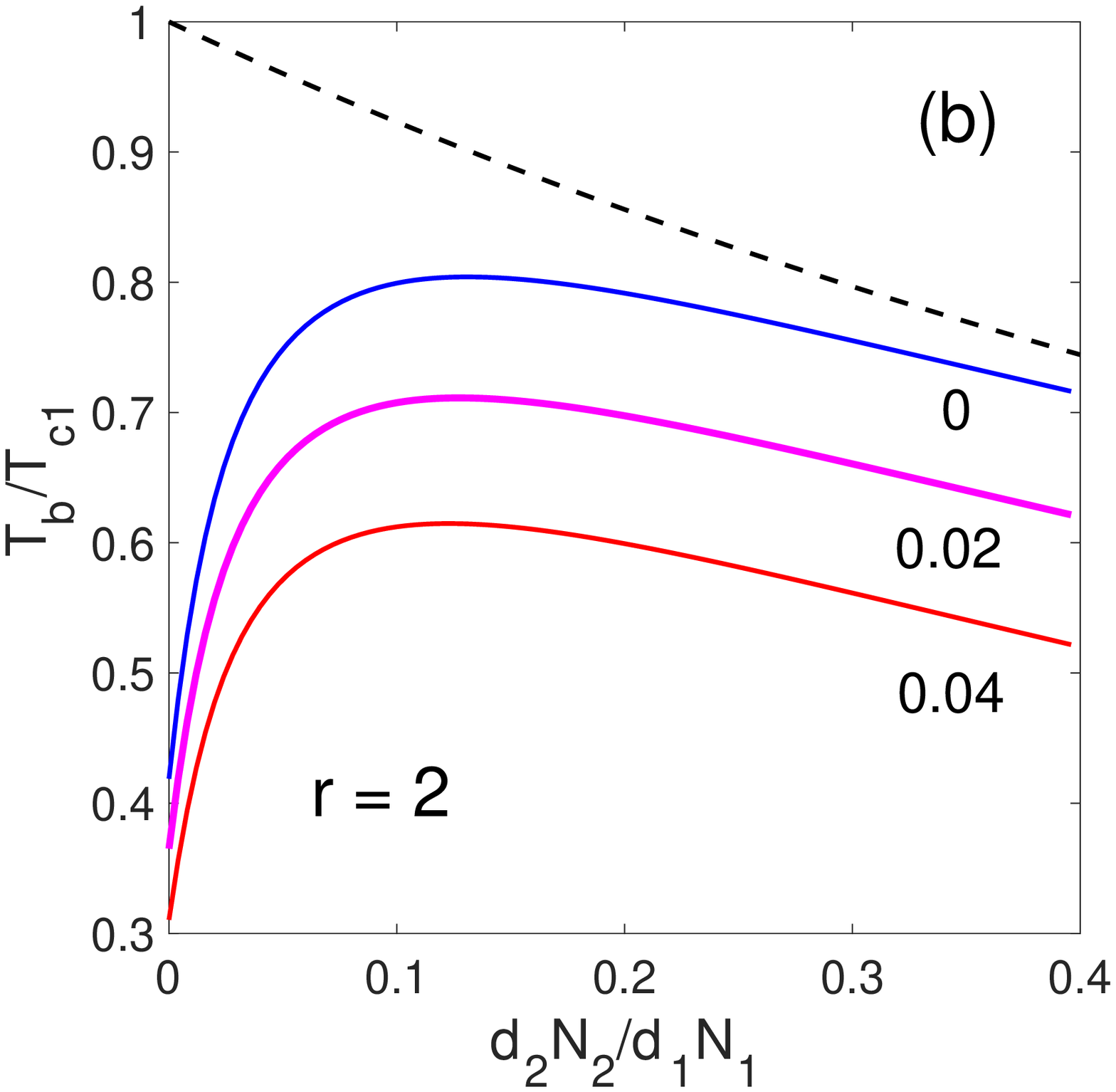}
   \caption{ BKT transition temperature $T_b(d_2)$ 
calculated from Eqs. (7), (27) and (47) at $D_2=50D_1$, $\lambda_1=0.7$, $\lambda_2=0.2$, $\Omega_2=2\Omega_1$ for different values of the DOS broadening parameter $\Gamma/2\pi T_{c0}$  and the resistance ratios $r=0,8$ (a) and $r=2$ (b). The dashed line shows the proximity effect-limited $T_{c0}$ of the bilayer in the absence of the BKT fluctuations.
   }\label{fig6}
\end{figure}

The DOS broadening reduces both $T_{c0}$ and the BKT transition temperature. For a single film, the DOS broadening does not change qualitative the dependence of $T_b$ on $r$ except for the overall reduction of $T_b(r)$ as shown in Fig.  \ref{fig1}. The effect of DOS broadening on the nonmonotonic dependence of $T_b(\alpha)$ in a bilayer with $R_B=0$ and $D_2\gg D_1$ is shown in Fig. \ref{fig6}, where $T_b(\alpha)$ was calculated from Eqs. (\ref{del}), (\ref{gap0}) and (\ref{bkt}). Here $T_b(\alpha)$ also  decreases as the broadening parameter $\Gamma/2\pi T_{c1}$ increases. This may be relevant to experiments \cite{dynes2} in which a nonmonotonic resistive transition temperature as a function of the overlayer thickness in Pb films was observed along with a reduction of $T_c$ and the DOS broadening. 

As the contact resistance increases, the proximity effect suppression of $T_{c0}$ diminishes. At the same time, a significant $R_B$ with $\beta \gtrsim 1$ tends to decouple the layers 1 and 2, 
suppressing the increase of the phase stiffness by the overlayer. The effect of these opposite trends on $T_b$ can be calculated by solving Eqs. (\ref{tb}), (\ref{t12}), (\ref{tt1}) and (\ref{kin}) numerically.  
At $\beta \gg 1$ the superfluid density caused by the proximity effect in the N overlayer is strongly reduced, and  
$T_b$ of a bilayer becomes limited by the induced weak superconductivity in the N layer, even if $\sigma_2\gg \sigma_1$.

\section{Finite size effects} 

Finite size effects can be essential is thin film bridges where, in addition to the BKT vortex unbinding, the resistive transition is affected by thermally-activated hopping 
of single vortices across the bridge and proliferation of fractional vortices in weakly-coupled bilayers. 

\subsection{Thermally-activated vortex hopping}
 
Dynamics of vortex hopping is determined by the local energy $U(u)$ of the vortex as a function of its position $u$ across the bridge. 
A vortex in a thin film strip of width $w<\Lambda$ produces circulating superflow with the normal components $Q_x(0,y)=Q_x(w,y)$ vanishing at the edges, and 
$\textbf{Q}(x,y)$ decreasesing exponentially over the length $w/\pi$ along the bridge \cite{film,ahm} (see Appendix \ref{ap3}). The energy barrier $U(u)$ in a strongly-coupled 
bilayer can be calculated in the same way as for a single film \cite{gv}, except that the vortex energy scale $\epsilon_0$ is now determined by the composite parameters defined by Eqs. (\ref{Sin}) and (\ref{del}):   
 \begin{equation}        %1
    U(u)=\zeta\epsilon_0\ln[(w/\pi\tilde{\xi})\sin(\pi u/w)],
    \label{u0}
    \end{equation}
where $\tilde{\xi}=C\xi$ is an effective coherence length, $C\approx 0.34$ accounts for the core energy \cite{gv}. 
The coherence length $\xi$ and the viscous drag coefficient $\eta$ of a vortex in a bilayer at $T\approx T_c$ were  evaluated in Appendix B:
\begin{gather}
\xi=\left[\frac{\pi\hbar D}{8(T_{c0}-T)} \right]^{1/2}, 
\label{xef}
\\
D=\frac{d_1N_1D_1+d_2N_2D_2}{d_1N_1+d_2N_2}.
\label{def}
\\
\eta=\frac{\phi_0^2}{2\pi\xi^2R_\square}=8\hbar(d_1N_1+d_2N_2) (T_{c0}-T).
\label{eta}
\end{gather}
Here the vortex core size $\xi$ defined by the composite diffusivity $D$ increases as the overlayer thickness increases, but  
the viscosity $\eta$, which takes into account dissipation in the vortex core in both layers, 
turns out to be independent of $\sigma_1$ and $\sigma_2$. The latter results from the fact that the diffusivity $D$ cancels out in the product $\xi^2R_\square$ 
in Eq. (\ref{eta}), thus $\eta$ in the Bardeen-Stephen model \cite{kopnin} becomes independent of the mean free paths.  

A solution of the Fokker-Planck equation for thermally-activated vortex hopping over the barrier
$U(x)$ gives the following votage-current (V-I) characteristics \cite{gv}  
    \begin{equation}
    V=\frac{2{\cal R}_nI(z-1)}{s\Gamma(z+1)}\left[\frac{2\pi\tilde{\xi}}{w}\right]^z\!
    \biggl|\Gamma\biggl(1+\frac{z}{2}+is\biggr)\biggr|^2\!\sinh\pi s,
    \label{exact}
    \end{equation}
where ${\cal R}_n=L/w(d_1\sigma_1+d_2\sigma_2)$ is the total normal state resistance, $z=\epsilon_0/T$, $s=\phi_0I/2\pi T$, and $\Gamma(x)$ is the gamma function.  At small currents, $s\ll 1$, Eq.
(\ref{exact}) yields the ohmic $V={\cal R} I$, where  
    \begin{equation}
    \frac{{\cal R}}{{\cal R}_n}=\frac{2\pi^{3/2}z\Gamma(z/2)}{
    \Gamma[(z-1)/2]}\left(\frac{\pi\tilde{\xi}}{w}\right)^z.
    \label{rv}
    \end{equation}
These formulas are applicable at $T<T_b$, that is, $z>2$. If $z\gg 1$ the vortex ohmic resistance 
${\cal R}\simeq\sqrt{2}{\cal R}_n(\pi z)^{3/2}(\pi\tilde{\xi}/w)^z\ll {\cal R}_n$ depends strongly on $w$. As $I$ increases 
the V-I characteristics at $s>1$ becomes nonlinear, $V\propto I^{z+1}$, and independent of the bridge width\cite{gv}. 

\begin{figure}[tb]
   	\includegraphics[width=8.5cm]{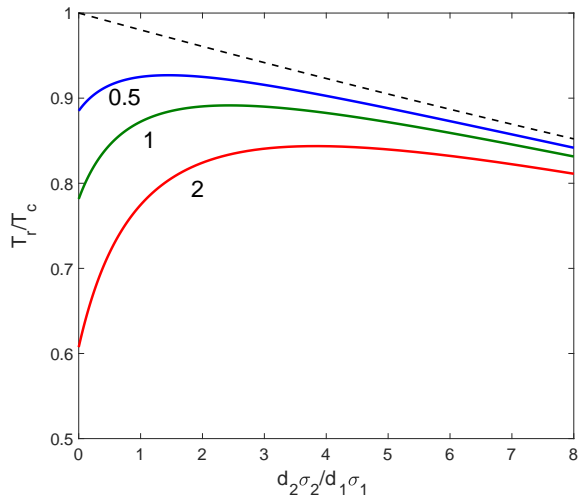}
   	\caption{Temperature of the resistive transition in the NS bilayer as a function of the overlayer thickness calculated from Eqs. (\ref{rv}) at the resistance criterion ${\cal R}_v=0.1{\cal R}$, 
   	$\lambda_1=0.5$, $D_2=100D_1$, $w=10\xi_1$, and different values of $r=8R/\pi \zeta R_0$: 0.5, 1, 2. The dashed line shows the mean-field $T_c(d_2)$.}
   	\label{fig7}
   	\end{figure}

The resistive transition temperature $T_r(I,w)$ calculated from Eqs. (\ref{xef})-(\ref{exact})     
depends not only on the conductivities and thicknesses of the layers but also on the width of the bridge and the electric field or resistance criterion at which $T_r$ is defined in transport measurements.   
For instance, Fig. \ref{fig7} shows $T_r(d_2)$ calculated from Eqs. (\ref{rv}) for the resistance criterion ${\cal R}=0.1{\cal R}_n$, $w=10\xi_1$, $D_2=100D_1$, 
and different ratios $r=8R/\pi\zeta R_0$. Here $z=(2p/r)\tanh p$ depends on the parameter $p=\Delta/2T_r$ which is obtained from Eq. (\ref{rv}) for a particular ${\cal R}_v$ and then used to calculate $T_r$  in Eq. (\ref{ktb}). The so-obtained dependence $T_r(d_2)$ for a single-vortex hopping appears similar 
to that of $T_b(d_2)$ calculated in the previous sections, the nonmonotonic dependence of $T_r(d_2)$ becomes more pronounced if the resistance criterion is chosen 
at a fixed ratio ${\cal R/R}_1$, where ${\cal R}_1=L/\sigma_1d_1w$.   

\begin{figure}[tb]
   	\includegraphics[width=11cm,trim={45mm 65mm 20mm 60mm},clip]{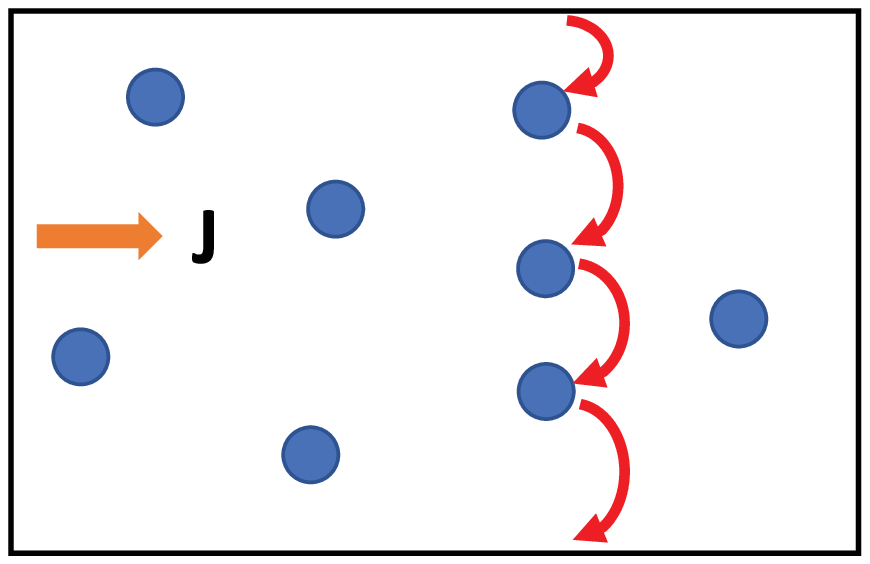}
   	\includegraphics[width=8cm]{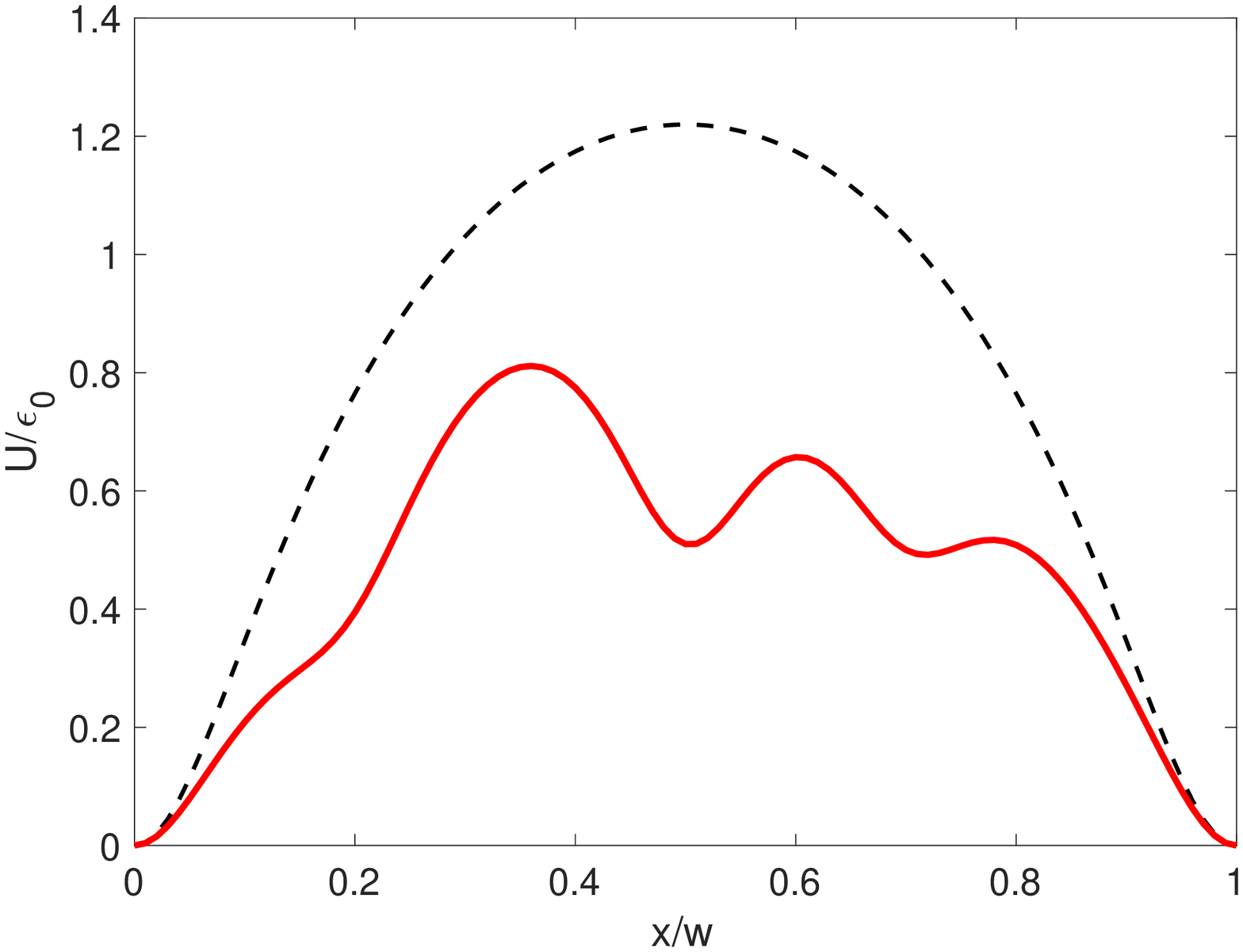}
   	\caption{Top: Thermally-activated hopping of the vortex along a chain of pinning centers shown as blue regions. Bottom: Sketch of the local energy of the vortex $U(x)$. The dashed line shows $U(x)$ in a uniform bridge  	calculated from Eq. (\ref{u0}) at $w=10\tilde{\xi}$. The solid line shows $U(x)$ given by Eq. (\ref{u0}) plus the pinning potential modeled by three Lorentzian wells, $U_p(x)=-\sum_i U_i\xi^2/[(x-x_i)^2+\xi^2]$ with 			$U_i=(0.3, 0.6, 0.4)\epsilon_0$ at $x_i=(0.2, 0.5,0,7)w$. The London core singularities at $x=0$ and $x=w$ were regularized to provide zero vortex energy at the edges, $U(0)=U(w)=0$.}
   	\label{fig8}
   	\end{figure}

These calculations of $V(I)$ and $T_r$ were based on Eq. (\ref{u0}) for the energy of a single vortex in a uniform bridge with no materials defects in the bulk and perfect film edges. This 
model is an idealization of a more realistic situation in which a bridge has materials defects at the edges and in the bulk, as depicted in Fig. \ref{fig8}. Defects such as nonsuperconducting second phase precipitates, 
grain boundaries or variation of the film thickness can pin vortices and   
lower local activation barriers, resulting in preferential hopping of vortices along chains of defects, as shown in Fig. \ref{fig8}.  Such behavior of vortices was 
recently observed in Pb films by SQUID on tip scanning microscopy \cite{embon}.

Pinning centers can facilitate thermally-activated vortex hopping and reduce $T_r$ as compared to a uniform bridge.  
However, a proximity coupled conductive overlayer can nearly restore $T_r$ back to $T_{c0}$ by increasing the vortex energy scale $\epsilon_0$ and by weakening the effect of pinning potential on vortex hopping.  Indeed, if pinning centers are in the S layer, deposition of the N overlayer would increase $\epsilon_0$ and vortex energy barriers without affecting the pinning energy. As a result, 
the nonmonotonic dependence of $T_r(d_2)$ becomes more pronounced because pinning mostly increases the dip in $T_r$ at $d_2=0$ while causing only a small correction to $T_r$ at larger $d_2$ for which the effect of overlayer becomes dominant.  
   
   \subsection{Partial vortices}
   
As was mentioned in Sect III, partial vortices may occur in a weakly-coupled bilayer with small Josephson current density $J_c$ across the interface between the layers 1 and 2. 
Fractional vortices have been investigated theoretically \cite{fv1} and observed in bilayers \cite{fv2}. Partial vortices could contribute to the resistive transition in 
short bilayer bridges $L<L_c$ at temperatures close to $T_{c1}$ of layer 1 for which the condition (\ref{lcc}) is satisfied. In this case the layers 1 and 2 become 
phase-unlocked so that the overlayer does not increase the kinetic energy of superflow around 
a vortex but produces a Josephson energy proportional to the area of the bridge. 
 
The energy of a perpendicular vortex in the granular film 1 can be reduced by weak intergranular contacts, but it does not affect Eq. (\ref{lcc}) which defines the condition under which fractional vortices can appear in both granular and nongranular bilayer. The above results are applicable for layers much thinner than the London penetration depth, 
$\lambda_L$ so that the layer 2 is transparent to the magnetic field produced by the vortex in the layer 1.  If $d_2>\lambda_L$ a thick overlayer 
traps the vortex magnetic field and spreads it along the interface between the layers 1 and 2. This increases the magnetic energy of the vortex and the BKT transition 
temperature \cite{magtune}. Such effect would be most pronounced in a thin film sandwiched between two massive superconductors.  

A different mechanism of mitigation of vortex fluctuations occurs if a disconnected N overlayer is spaced by a wide gap of width $d_i$ from the superconducting layer 1. It was observed that a 30 nm thick Au 
overlayer separated by 16 nm gap from 3 nm thick MoGe film slightly increases the temperature of the resistive transition \cite{kap}. This effect was associated with additional dissipation caused by eddy currents induced by a moving vortex in the metallic overlayer, mitigating quantum tunneling of vortices \cite{finkels}. Here we consider the influence of a remote N overlayer  
on thermally-activating hopping of vortices. This process is controlled by the vortex drag coefficient $\eta$ which was calculated in Appendix \ref{ap4}: 
\begin{equation}
\eta=\frac{\phi_0^2d_1}{2\pi\xi_1^2\rho_1}+\frac{(\ln 4-1)\phi_0^2d_2}{32\pi\Lambda_1^2\rho_2},
\label{etta}
\end{equation}   
where the first term in the right hand side is the Bardeen-Stephen drag coefficient for a vortex in the S film, and the second term is the inductive  
drag coefficient $\eta_2$ due to the metallic overlayer. Here $\eta_2$ is consistent up to a numerical factor $\sim 1$ with the result of Ref. (\onlinecite{finkels}) obtained in the limit of $d_i=0$. 
As shown in Appendix \ref{ap4}, $\eta_2$ turns out to be independent of the gap width $d_i$ as long as $d_i+d_2\ll \min (w,\Lambda)$. Although $\eta_2$ 
appears similar to $\eta_1$ with the replacement $\rho_2\to\rho_1$ and $\xi\to\Lambda$, the inductive heating in the overlayer actually occurs in a small region of radius $\sim d_i+d_1\ll w$. 
Here the factor $\Lambda^{-2}$ in $\eta_2$ does not result from magnetic screening but comes from the magnitude of vortex sheet current in the moving Pearl vortex \cite{pearl} 
which induces eddy currents in the overlayer.     

Very thin films have $\Lambda^2\gg \xi^2$ so $\eta_2$ is generally much smaller than $\eta_1$, even for highly conductive overlayers with $\rho_2\ll \rho_1$.  
The ratio of the inductive and viscous drag coefficients is:
\begin{equation}
\frac{\eta_{2}}{\eta_{1}}\simeq \frac{d_{1}d_{2}\rho_1}{40\kappa^{4}\xi^{2}\rho_2},
\label{condit}
\end{equation}
where $\kappa=\lambda_{L}/\xi$ is the GL parameter. For the amorphous MoGe films with $\kappa\sim100$, $\xi\simeq 25(1-T/T_{c})^{-1/2}$ nm,  $\rho_{1}\simeq 200\mu\Omega$cm \cite{stanf},  $d_{1}=3$ nm, and the Au overlayer with $d_{2}=40$ nm and  $\rho_{2}=22n\Omega$ cm investigated in Ref. \onlinecite{kap}, Eq. (\ref{condit}) gives $\eta_2/\eta_1 \sim 10^{-6}$.  

\section{Discussion}

The resistive transition temperature in thin superconducting films can be tuned by overlayers which  
ameliorate pairbreaking fluctuation of vortices and shift $T_r$ back to the mean-field $T_c$. Revealing  
the actual $T_c$ of a new 2D superconductor could be done using: 
1. S-I-S$'$ trilayers in which a known higher-$T_c$ superconductor S$'$ is deposited onto a new superconductor S separated by a thin dielectric layer, 2. A bilayer in which a lower-$T_c$ superconductor or normal overlayer with high carrier density or normal state conductivity is deposited onto a superconducting film.  3. Metallic or superconducting overlayers which are capacitively or inductively coupled with the main superconducting film. The first two approaches rely on static mechanisms which increase energies of vortices. The third approach is based on dynamic mechanisms which affect quantum fluctuations and increase the vortex drag, making vortices less mobile. 

\begin{enumerate}[leftmargin=4mm]
\item
S-I-S$'$ trilayers could be used to reveal $T_c$ of new materials (for instance, FeSe single layers) using high-$T_c$ overlayers. In this case the current is injected into the $S$ layer and spreads along both layers over the Josephson length $L_J$ which determines the scale of current re-distribution. The solution for the phase difference $\chi(x)=\chi_2-\chi_1$ obtained in Appendix \ref{ap4} is:
\begin{gather}
\tan\frac{\chi}{4}=\frac{I e^{-x/L_J}}{I_b+\sqrt{I_b^2-I^2}},
\label{chi} \\
I_g=\frac{2d_1g_1}{L_J},\qquad L_J = \left[\frac{d_1d_2g_1g_2}{(d_1g_1+d_2g_2)J_c}\right]^{1/2}.
\label{LJ}
\end{gather}
Here $J_c$ is the Josephson current density through the interface, and the phase conductivities $g_1$ and $g_2$ define the current densities $\textbf{J}_1=g_1\nabla\chi_1$ and $\textbf{J}_2=g_2\nabla\chi_2$ in the layer 1 and 2 due to the respective phase gradients $\nabla\chi_1$ and $\nabla\chi_2$. For dirty s-wave superconductors, $g_i=(\pi\Delta_i\sigma_i/2e)\tanh (\Delta_i/2T)$, $i=1,2$. At $I>I_b$ the current injected into the layer 1 generates interlayer phase slips \cite{mlphase}. Therefore, the S$'$ layer does not short circuit the $S$ layer if $I<I_b$, and the length of the bridge is shorter than $L_J$. 

A higher-$T_c$ overlayer increases the energy barriers for the BKT proliferation or thermally-activated hopping of perpendicular vortices, depending on the overlayer thickness $d_2$, as illustrated in Fig. \ref{fig2}. At small $d_2<d_{2c}$, the overlayer increases the phase stiffness and the energies of complete vortices threading both layers, so that $T_r(d_2)$ increases with $d_2$ up to the critical thickness $d_{2c}$ defined by Eq. (\ref{alk}).  At $d_2>d_{2c}$ partial vortices in layer 1 become more energetically favorable and the overlayer increases the energy of the vortex by the amount of the Josephson energy proportional to the area of the bridge, so that $T_r$ becomes independent of $d_2$. The maximum value of $T_r$ at $d_2>d_{2c}$ can be reached by changing the bridge dimensions and the interlayer $J_c$. 

\item
The resistive transition temperature $T_r$ can be increased in a bilayer with a proximity-coupled overlayer which can be either normal or superconducting. Here partial vortices are not energetically favorable, but the overlayer increases the total sheet superfluid density and thus the energy of complete vortices while decreasing the mean-field $T_c$ due to the proximity effect. As was shown above, the interplay of these trends yields a nonmonotonic dependence of $T_r$ and the BKT transition temperature on the overlayer thickness. 

The maximum $T_r$ close to the mean-field $T_c$ could be reached by depositing a thin normal layer with $d_2 \ll d_1$, where the optimum thickness $d_{2m}$ estimated by Eq. (\ref{d2m}) turns out to be independent of $\sigma_1$ if $\sigma_2\gg\sigma_1$. This condition is satisfied for good metals such as Ag, Cu or Au with $\sigma_2\sim (10^3-10^4)\sigma_1$ as compared to typical values of $\sigma_1$ for cuprates, pnictides or amorphous low-$T_c$ monolayers. The proximity-effect reduction of $T_c$ can be ameliorated by the contact resistance between the layers 1 and 2, as shown in Fig. \ref{fig3}. In turn, the contact resistance can be effectively tuned by heat treatment which can change $R_B$ by several orders of magnitude as, for example, was shown for the YBCO-Ag interface~ \cite{int1,int2}. 
  
\item
Fluctuations in a 2D superconductor can be tuned by its inductive or capacitive coupling with a remote normal or superconducting film. This effect was observed  on planar arrays of Al Josephson junctions  \cite{jjatunee} and MoGe films \cite{kap}. Theoretical explanations invoked the ideas of remote gates providing tunable dissipative environment affecting quantum fluctuations and tunneling of vortices in a superconductor \cite{jjatunet, finkels}. For thermally-activated dynamics of vortices considered in this paper, a remote gate causes additional vortex drag due to eddy currents induced in a metallic overlayer \cite{finkels}. However, the inductive contribution to the vortex drag coefficient $\eta_2$ in Eqs. (\ref{etta}) and (\ref{condit}) turns out to much smaller that the conventional Bardeen-Stephen viscous drag in the superconducting film, particularly in the extreme 2D limit, $\Lambda^2/\xi^2\to\infty$.  Therefore, despite the proximity effect reduction of $T_c$, the increase of $T_r$ by direct contact of the S film with a thin, highly conducting normal layer appears far more effective than increasing the vortex drag by inductive coupling.

\end{enumerate}

The approach of this work is based on the conventional Usadel equations assuming that the pairing constants, normal densities of states and phonon frequencies are independent of the layer thicknesses. This model takes into account neither surface scattering nor interface superconductivity caused by localized phonon modes and changes the pairing constants and DOS at the interface. For instance, a highly conductive overlayer can improve electron screening in the S layer, weakening the Coulomb repulsion and enhancing the Cooper pairing \cite{coulomb1,coulomb2}. In this case one would expect that the mean-field $T_{c0}(d_2)$ increases as $d_2$ increases, levels off as $d_2$ exceeds the Thomas-Fermi screening length $l_{TF}$ and then decreases at larger $d_2$ due to the proximity effect.  However, the small values of   
$l_{TF}=0.5-0.6~\AA$ for Pb, Cu, Ag and Au \cite{Ash} indicate that the effect of screening on $T_c$ becomes independent of the overlayer thickness at $d_2 \gtrsim 1 \AA$. In this case screening may not explain the non-monotonic dependence of $T_r(d_2)$ with maxima at $2-4~\AA\gg l_{TF}$ observed on Bi-(Au, Ag), Ga-Ag and Pb-Ag bilayers \cite{bi1,bi2,ga,pb}. The maxima in $T_r(d_2)$ at $d_2\gg l_{TF}$ readily follow from the vortex mechanism suggested in this work.     

Overlayers can be used to tune the BKT transition and reveal the effect of different materials parameters, particularly, inhomogeneities \cite{disordbkt,broad}, DOS broadening and surface and interface scattering. Given the significant DOS broadening observed by tunneling experiments on ultra thin films \cite{pb,tun,blstm1,blstm2,stmPb}, the pairbreaking DOS broadening effects can contribute to the observed reduction of both $T_c$ and  $T_b$. Since the DOS broadening affects $T_c$ and $T_b$ differently, it cannot be just taken into account by substituting the observed $T_c$ into Eq. (\ref{tb}) to infer $T_b$ from the experiment. 

The BKT transition temperature depends on the factor $\zeta$ affected by multiple mechanisms contributing to the renormalization of the superfluid density and electron diffusivity by strong electron-phonon coupling \cite{lam1,lam2,lam3}, fluctuations and weak localization effects \cite{bktm}. Moreover, $\zeta$ can be affected by such uncertain materials factors as inhomogeneities of $T_c$, defects which pin vortices, crystalline granularity, DOS broadening, surface scattering and finite size effects. Thus, the actual evaluation of $T_b(d_2)$ controlled by the resistance ratio $r=8R_0/\pi\zeta R$ can only be done if $\zeta$ is regarded as a material parameter which could be expressed via the observed $T_b$ of a bare film at $d_2=0$. This paper focuses on qualitative effects of the overlayer on the resistive transition temperature which was quantified by either $T_b(d_2)$ or $T_r(d_2)$ for single-vortex hopping. It turned out that both $T_b(d_2)$ and $T_r(d_2)$ have similar dependencies on $d_2$, so the main conclusion about the mitigation of vortex fluctuations by overlayers is not that sensitive to the resistance criterion  for $T_r$.  Other factors such the effect of the vortex core on the BKT transition in a bilayer where the core size given by Eq. (\ref{xi}) depends on $d_2$ and can be much larger than $\xi_1$ in the S film, 
deserves a more detailed investigation.       

\acknowledgments

This work was supported by AFOSR under grant FA9550-17-1-0196.

\appendix 

\section{Critical temperature of a bilayer} 
\label{ap1}

In the Cooper limit $\theta_{1,2}(x)$ are nearly uniform across the layers, so that the quadratic expansions can be used:
\begin{eqnarray}
\theta_1(x)=\theta_1-C_1(x+d_1)^2,
\label{teta1} \\
\theta_2(x)=\theta_2+C_2(x-d_2)^2.
\label{teta2}
\end{eqnarray}
Solution of Eqs. (\ref{u1})-(\ref{bc2}) at $C_1d_1^{2}\ll1$ and $C_2 d_2^{2}\ll1$ is
\begin{gather}
C_1 D_1=\Delta_1\cos\theta_1-\omega_1\sin\theta_1,
\label{c1} \\
C_2 D_2=\omega_2\sin\theta_2-\Delta_2\cos\theta_2,
\label{c2} \\
C_1 d_1\sigma_1=C_2 d_2\sigma_2,
\label{c12} \\
C_2d_2\sigma_2=R_B^{-1}(\sin\theta_1\cos\theta_2-\cos\theta_1\sin\theta_2),
\label{c21}
\end{gather}
where $\omega_{1,2}=\omega+\Gamma_{1,2}$. Solving for $C_1$ and $C_2$ yields Eq. (23)-(25). At negligible contact resistance $R_B\to 0$,  Eqs. (\ref{c1})-(\ref{c21}) give $\theta_1=\theta_2\equiv \theta$, and
\begin{gather}
\sin\theta=\frac{\Delta}{\sqrt{(\omega+\Gamma)^2+\Delta^2}}, 
\label{ssin} \\
\Delta=\frac{\Delta_1+\alpha\Delta_2}{1+\alpha}, \quad \Gamma=\frac{\Gamma_1+\alpha\Gamma_2}{1+\alpha},\quad \alpha=\frac{d_2N_2}{d_1N_1}.
\label{param}  
\end{gather}
The equations for $\Delta_1$ and $\Delta_2$ become
\begin{gather}
\Delta_1=2\pi T\lambda_1\sum_{\omega>0}^{\Omega_1}\frac{\Delta}{\sqrt{(\omega+\Gamma)^{2}+\Delta^{2}}}
\label{d1} \\
\Delta_2=2\pi T\lambda_2\sum_{\omega>0}^{\Omega_2}\frac{\Delta}{\sqrt{(\omega+\Gamma)^{2}+\Delta^{2}}}
\label{d2}
\end{gather}
Multiplying Eq. (\ref{d1}) by $1/(1+\alpha)$ and
Eq. (\ref{d2}) by $\alpha/(1+\alpha)$ and adding them gives a single equation for $\Delta$:
\begin{equation}
\!\!1=\sum_{\omega>0}^{\Omega_1}\frac{2\pi T\tilde{\lambda}_1}{\sqrt{(\omega+\Gamma)^2+\Delta^2}}+\sum_{\omega>0}^{\Omega_2}\frac{2\pi T\tilde{\lambda}_2}{\sqrt{(\omega+\Gamma)^2+\Delta^2}},
\label{dela}
\end{equation}
where 
\begin{equation}
\tilde{\lambda}_1=\frac{\lambda_1}{1+\alpha},\qquad \tilde{\lambda}_2=\frac{\lambda_2\alpha}{1+\alpha}.
\label{tlam}
\end{equation}
Taking the limit of $\Delta\to 0$ yields the equation $T_c$:
\begin{equation}
1=\sum_{n=0}^{\Omega_1/2\pi T_{c}}\frac{\tilde{\lambda}_1}{n+\frac{1}{2}+\gamma}+\sum_{n=0}^{\Omega_2/2\pi T_{c}}\frac{\tilde{\lambda}_2}{n+\frac{1}{2}+\gamma},
\label{ttt}
\end{equation}
where $\gamma=\Gamma/2\pi T$.
The summation in Eq. (\ref{ttt}) is not well defined because the hard cutoffs ${\cal N}_{1,2}=\Omega_{1,2}/2\pi T$ are not necessarily integer. Taking only integer parts of ${\cal N}_{1,2}$ 
in numerical calculations can produce spurious contributions in $T_c$, particularly if ${\cal N}_{1,2}$ are not very large for real materials.    
This issue can be addressed by 
inserting the bell-shape functions $S_{1,2}(n)={\cal N}_{1,2}^2/[(n+1/2)^2+{\cal N}_{1,2}^2]$ and extending the summation over $n$ to infinity. Then Eq. (\ref{ttt}) becomes
\begin{equation}
1=\sum_{n=0}^\infty\frac{\tilde{\lambda}_1S_1+\tilde{\lambda}_2S_2}{n+\frac{1}{2}+\gamma},
\label{tcw}
\end{equation}
The summation is done using:
\begin{gather}
I=\sum_{n=0}^\infty\frac{{\cal N}^2}{(n+\frac{1}{2}+\gamma)[(n+\frac{1}{2})^2+{\cal N}^2]}=
\nonumber \\
\!\!\frac{{\cal N}^2}{{\cal N}^2+\gamma^2}\!\left[\mbox{Re}\psi\!\left(\frac{1}{2}+i{\cal N} \right)-\psi\!\left(\frac{1}{2}+\gamma\right)+ \frac{\pi\gamma}{2{\cal N}}\tanh\pi{\cal N}\right]\!.
\label{sum}
\end{gather}
If ${\cal N}\gg \max (\gamma,1)$, $\mbox{Re}\psi(\frac{1}{2}+i{\cal N})\simeq \ln{\cal N}$ so that
\begin{equation}
I=\ln(4\gamma_E {\cal N})-U(\gamma),
\label{I} 
\end{equation}
where $U(\gamma)$ is defined by Eq. (\ref{U}), and $\psi(\frac{1}{2})=-\ln(4\gamma_E)$. At $\gamma=0$, Eqs. (\ref{tcw}) and (\ref{I}) reproduce the well-known $T_{c0}$ of a bilayer in the Cooper limit \cite{cooper}: 
\begin{equation}
T_{c0}=\frac{2\gamma_E}{\pi}\Omega_1^{1-a}\Omega_2^{a}e^{-1/\lambda},
\label{tcoo}
\end{equation}
where
\begin{equation}
\lambda=\frac{\lambda_1+\alpha\lambda_2}{1+\alpha},
\qquad
a=\frac{\alpha\lambda_2}{\lambda_1+\alpha\lambda_2}.
\label{zet}
\end{equation}

If $\Gamma_1$ and $\Gamma_2$ are essential, $T_c$ is determined by Eq. (\ref{tcw}) which can be recast in the form:
\begin{equation}
\!\!\!1-\sum_{n=0}^\infty\frac{\tilde{\lambda}_1S_1+\tilde{\lambda}_2S_2}{n+\frac{1}{2}}=\lambda\sum_{n=0}^\infty\!\left[\frac{1}{n+\frac{1}{2}+\gamma}-\frac{1}{n+\frac{1}{2}}\right].
\label{tttt}
\end{equation}
Here the second term in the left hand side was subtracted from both sides of Eq. (\ref{tcw}). The sum in the right hand side converges over $n\sim\gamma\ll{\cal N}_{1,2}$, so $S_{1,2}(n)$ were set to $1$, and $\lambda=\tilde{\lambda}_{1}+\tilde{\lambda}_2$ was used. Summing up in Eq. (\ref{tttt}) using Eqs. (\ref{I})-(\ref{zet}) yields Eq. (\ref{tc1}). 

If the interface resistance cannot be neglected, Eqs. (\ref{t12})-(\ref{am}) for $\theta_1$ and $\theta_2$ can only be solved numerically. 
A general equation for $T_c$ can be obtained by linearizing Eqs. (\ref{t12}) and (\ref{tt1}) with respect to small $\theta_1$ and $\theta_2$:  
\begin{gather}
\theta_{1}=\frac{\Delta_{1}(1+\alpha\beta\omega_{2})+\alpha\Delta_{2}}{(1+\alpha\beta\omega_{2})\omega_1+\alpha\omega_{2}},
\label{tet1} \\
\theta_{2}=\frac{\Delta_{1}+\alpha(1+\beta\omega_{1})\Delta_{2}}{\alpha (1+\beta\omega_1)\omega_2+\omega_{1}}.
\label{tet2}
\end{gather}
Substituting Eqs. (\ref{tet1}) and (\ref{tet2}) into the linearized Eq. (\ref{gap12}) and solving the resulting system of linear equations for 
$\Delta_1$ and $\Delta_2$ yields the following equation for $T_c$:
\begin{equation}
(1-\lambda_1R_{11})(1-\lambda_2R_{22})-\lambda_1\lambda_2R_{12}R_{21}=0,
\label{genTc}
\end{equation}
where
\begin{gather}
R_{11}=2\pi T_c\sum_{\omega>0}^\infty\frac{(1+\alpha\beta\omega_{2})S_1(\omega)}{(1+\alpha\beta\omega_2)\omega_1+\alpha\omega_2},
\label{r11} \\
 R_{22}=2\pi T_c\sum_{\omega>0}^\infty\frac{\alpha(1+\beta\omega_{1})S_2(\omega)}{\alpha(1+\beta\omega_1)\omega_2+\omega_1},
\label{r22} \\
R_{12}=2\pi T_c\sum_{\omega>0}^\infty\frac{\alpha S_1(\omega)}{(1+\alpha\beta\omega_2)\omega_1+\alpha\omega_2}, 
\label{r12} \\
R_{21}=2\pi T_c\sum_{\omega>0}^\infty\frac{S_2(\omega)}{\alpha(1+\beta\omega_1)\omega_2+\omega_1}.
\label{r21}
\end{gather}
Equations (\ref{genTc})-(\ref{r21}), which contain rapidly converging sums, are rather suitable for numerical calculations of $T_c$ depending on the multitude of materials parameters  
$\lambda_{1,2}, d_{1,2}, N_{1,2}, \Gamma_{1,2}, R_B$.

For a normal overlayer with $\lambda_2=0$, the equation for $T_c$ takes the form (\ref{tcc}). If $\Gamma_1=\Gamma_2=0$, this equation can be reduced to:
\begin{equation}
\frac{1}{\lambda_1}=\frac{1}{1+\alpha}\sum_{n=0}^\infty\left[\frac{1}{n_1}+\frac{\alpha}{n_1+{\cal M}} \right]\frac{{\cal N}^2}{n_1^2+{\cal N}^2}
\label{tc10}
\end{equation}  
where $n_1=n+1/2$, ${\cal N}=\Omega_1/2\pi T$, and ${\cal M}=(1+\alpha)/2\pi \alpha\beta T$. Summation in Eq. (\ref{tc10}) can be done using Eq. (\ref{sum}). 
In the BCS limit ${\cal N}\gg 1$, one can use $\mbox{Re}\psi(1/2+i{\cal N})\to \ln {\cal N}$ so that Eq. (\ref{tc10}) becomes 
\begin{gather}
\frac{1+\alpha}{\lambda_{1}}=\ln\frac{2\gamma_{E}\Omega_1}{\pi T}+
\nonumber \\
\frac{\alpha {\cal N}^2}{{\cal N}^2+{\cal M}^2}\left[\ln\frac{2\gamma_{E}\Omega_1}{\pi T}+\frac{\pi{\cal M}}{2{\cal N}}-U({\cal M})\right].
\label{tc11}
\end{gather}
Using here $(1+\alpha)/\lambda_1=\ln(2\gamma_E\Omega_1/\pi T_{c0})$, where $T_{c0}=T_{c1}\exp(-\alpha/\lambda_1)$ is the critical temperature of a N-S bilayer with $R_B=0$, and 
substituting ${\cal N}=\Omega_1/2\pi T$ and ${\cal M}=(1+\alpha)/2\pi \alpha\beta T$ yields Eq. (\ref{Tcc}).
 
It is  instructive to compare $T_c$ described by Eq. (\ref{Tcc}) with $T_c$ obtained using the BCS hard cutoff at $\omega=\Omega_1$ in which case 
Eq. (\ref{tc10}) is truncated to
\begin{equation}
\frac{1}{\lambda_1}=\frac{1}{1+\alpha}\sum_{n=0}^{{\cal N}}\left[\frac{1}{n_1}+\frac{\alpha}{n_1+{\cal M}} \right] 
\label{tc12}
\end{equation}
Hence,
\begin{gather}
\frac{1+\alpha}{\lambda_1}=
\psi\left(\frac{3}{2}+{\cal N}\right)-\psi\left(\frac{1}{2}\right)+ 
\nonumber \\
\alpha\left[\psi\left(\frac{3}{2}+{\cal M}+{\cal N}\right)-\psi\left(\frac{1}{2}+{\cal M} \right)\right]
\label{tc13}
\end{gather}  
In the BCS limit ${\cal N}\gg 1$ Eq. (\ref{tc13}) yields the following equation  
which has been obtained previously \cite{fominov,gol}:
\begin{equation}
\!\!\!\!\ln\frac{T_c}{T_{c1}}=\frac{\alpha}{1+\alpha}\left[\ln\!\left(1+\frac{1+\alpha}{\alpha\beta\Omega_1}  \right)-U\!\left(\frac{1+\alpha}{2\pi\alpha\beta T_c} \right)\right]\!,
\label{tc14} 
\end{equation}
where the logarithmic term in the brackets which provides the correct limit $T_c\to T_{c0}(\alpha)$ at $R_B\to 0$ is only essential at small $\alpha$ and $\beta$ for which ${\cal M}\gg 1$.
Numerical solutions show that both $T_c(\alpha)$ described by Eq. (\ref{Tcc}) and (\ref{tc14}) have very similar dependencies on $\alpha$. For the case shown in Fig. \ref{fig3}, 
the largest difference ($\lesssim 9\%$) between $T_c(\alpha)$ calculated from Eq. (\ref{Tcc}) and (\ref{tc14}) occurs at $2\pi\beta T_{c1}=3$ and $\alpha\simeq 1$.

\section{Free energy, GL functional, vortex core energy and viscosity. }
\label{ap2}

The free energy of a dirty bilayer is given by 
\begin{gather}
F=\int (f_1+f_2)d^2{\bf r}, 
\label{F12} \\
f_l=\frac{\nu_l\Delta_l^{2}}{\lambda_l}+4\pi Tn_l\sum_{\omega>0}^{\Omega_l}\biggl[\omega(1-\cos\theta_l)-\Delta_l\sin\theta_l 
\nonumber \\
+\frac{D_l}{2}Q^2\sin^2\theta_l+\frac{D_l}{2}(\nabla\theta_l)^2\biggr],
\label{Fi}
\end{gather}
where $\nu_{i}=d_{i}N_{i}$, $\textbf{Q}=\nabla\chi +2\pi\textbf{A}/\phi_0$, $\chi_l(\textbf{r})$ is the phase of the order parameter, $\Psi_l(\textbf{r})=\Delta_l(\textbf{r})e^{i\chi_l(\textbf{r})}$, and $l=1,2$. For 
a strongly coupled bilayer with $\alpha\beta\Omega_{1,2}\ll 1$ and no DOS broadening, $\theta_1=\theta_2=\theta$ and Eq. (\ref{gap12}) yields 
\begin{gather}
\lambda_1\Delta_2=\lambda_2\Delta_1,\qquad \nu\Delta=\nu_1\Delta_1+\nu_2\Delta_2,
\label{param1}\\
\Delta_1=\frac{\lambda_1}{\lambda}\Delta,\qquad\Delta_2=\frac{\lambda_2}{\lambda}\Delta,
\label{param2} \\
\Delta=\frac{\nu_1\Delta_1+\nu_2\Delta_2}{\nu_1+\nu_2}, \qquad \lambda=\frac{\lambda_1\nu_1+\lambda_2\nu_2}{\nu_1+\nu_2},
\label{param3}
\end{gather}
where $\nu=\nu_1+\nu_2$. From Eqs. (\ref{param1})-(\ref{param3}), it follows,
\begin{equation}
\frac{\nu_1\Delta_1^{2}}{\lambda_1}+\frac{\nu_2\Delta_2^{2}}{\lambda_2}=\frac{\nu\Delta^{2}}{\lambda}.
\label{param4}
\end{equation}
Using Eqs. (\ref{param1})-(\ref{param4}) and Eqs. (\ref{tcoo})-(\ref{zet}) for $T_{c0}$, the free energy in Eqs. (\ref{F12})-(\ref{Fi}) for a phase-locked bilayer can be expressed in terms of a single order parameter $\Delta$, 
the combined sheet density of states $\nu=d_1N_1+d_2N_2$ and an effective diffusivity $D$:
\begin{gather}
F=\nu\int\biggl\{\Delta^{2}\ln\frac{T}{T_{c0}}+4\pi T\sum_{\omega>0}\biggl[\omega(1-\cos\theta)-\Delta\sin\theta
\nonumber \\
+\frac{\Delta^2}{2\omega}+\frac{D}{2}Q^2\sin^2\theta+\frac{D}{2}(\nabla\theta)^2\biggr]\biggr\}d^2{\bf r},
\label{ff} \\
D=\frac{\nu_1D_1+\nu_2D_2}{\nu_1+\nu_2}.
\label{D}
\end{gather}
Variation of $F$ with respect to $\delta\theta$ and $\delta\Delta$ results in the mean-field Usadel equations
\begin{gather}
2\omega\sin\theta+D(Q^2\sin\theta\cos\theta-\nabla^2\theta)=2\Delta\cos\theta,
\label{usa1} \\
\Delta\ln\frac{T}{T_{c0}}=2\pi T\sum_{\omega>0}\left(\sin\theta-\frac{\Delta}{\omega}\right).
\label{usa2}
\end{gather}

The GL equations are obtained by expanding Eq. (\ref{usa1}) in small gradients and powers of $\Delta$ at $T\approx T_c$:
\begin{equation}
\theta\approx\frac{\Delta}{\omega}+\frac{D}{2\omega^2}(\nabla^2\Delta-Q^2\Delta)-\frac{\Delta^3}{3\omega^3}.
\label{glus}
\end{equation}  
Substituting this into Eq. (\ref{ff}) and summing up over $\omega$ yields the GL functional (\ref{gl})-(\ref{c}).

A fluctuation contribution to the specific heat $\delta C(T)$ at $T>T_{c0}$ is obtained by expanding Eq. (\ref{ff}) to quadratic terms in the Fourier components $\Psi_k$:
\begin{equation}
\delta F=A\nu\int\frac{d^2{\bf k}}{(2\pi)^2}\biggl[\biggl(\ln\frac{T}{T_{c0}}+\frac{\hbar Dk^2}{8T_{c0}}\biggr)|\Psi_k|^2\biggr],
\label{dF}
\end{equation}
where $A$ is the bilayer area. The Gaussian fluctuation correction \cite{fluct} to the statistical sum $\delta Z=\int e^{-\delta F/T}D\Psi_k$ yields Eq. (\ref{dC}) for  
$\delta C=-T\partial^2\ln Z/\partial T^2$.

The condensation energy density $f_0$ of a uniform state is obtained by expressing $\Delta\ln (T/T_{c0})$ in Eq. (\ref{ff}) in terms of the 
$\omega$-sum from Eq. (\ref{usa2}):  
\begin{equation}
f_0=2\pi T\sum_{\omega>0}\frac{\nu\Delta^{4}}{\sqrt{\omega^{2}+\Delta^{2}}(\omega+\sqrt{\omega^{2}+\Delta^{2}})^{2}}.
\label{f0}
\end{equation}
Here $f_0=H_c^2/8\pi$ defines the thermodynamic critical field $H_c$ of a bilayer. 
At $T\approx T_{c0}$ the gap $\Delta$ in the denominator of Eq. (\ref{f0}) can be dropped giving
\begin{equation}
f_0=\frac{7\zeta(3)\nu\Delta^4}{16\pi^2 T_{c0}^2}.
\label{foc}
\end{equation}
The energy of the vortex core $\epsilon_c$ may be evaluated by writing the total energy of a vortex in the form:
\begin{equation}
\epsilon=\epsilon_0\ln\frac{L}{\xi}+2\pi \xi^2 f_0,
\label{eval}
\end{equation}  
where $\epsilon_0=(\pi^2/2)\hbar\nu D\Delta\tanh(\Delta/2T)$ follows from Eq. (\ref{epso}) at $\Gamma=0$, and the term $2\pi\xi^2 f_0$ accounts for the loss of condensation energy in a normal core of radius $\simeq \sqrt{2}\xi\,$\cite{clem}. The composite coherence length $\xi$ can be obtained by minimizing $\epsilon(\xi)$ with respect to $\xi$:
\begin{gather}
\epsilon=\epsilon_0\left(\ln\frac{L}{\xi}+\frac{1}{2}\right),\qquad \epsilon_c=\frac{\epsilon_0}{2},
\label{epev} \\
\xi=\left(\epsilon_0/4\pi f_0\right)^{1/2}.
\label{xie}
\end{gather} 
Using here $\epsilon_0=\pi^2\hbar D\nu \Delta^2/4T_{c0}$,  $\Delta^2(T)=8\pi^2 T_{c0}(T_{c0}-T)/7\zeta(3)$ and $f_0$ from Eq. (\ref{foc}) at $T\approx T_{c0}$ yields 
\begin{equation}
\xi=\left[\frac{\pi\hbar D}{8(T_{c0}-T)} \right]^{1/2}.
\label{xic}
\end{equation}
For a single film, Eq. (\ref{epev}) is consistent with $\epsilon_c\approx 0.497\epsilon_0$ obtained from numerical simulation of a single vortex \cite{hu}, 
and Eq. (\ref{xic}) reproduces the GL coherence length $\xi=\sqrt{c/|a|}$ with $a$ and $c$ given by Eqs. (\ref{gl})-(\ref{c}).  This qualitative analysis shows   
that both $\epsilon_c\propto (d_1\sigma_1+d_2\sigma_2)$ and the core radius $\xi\propto (d_1\sigma_1+d_2\sigma_2)^{1/2}$ of a vortex in a bilayer can be significantly  
increased by a conductive overlayer with $\sigma_2\gg \sigma_1d_1/\sigma_2$. 

The viscous drag coefficient $\eta$ of a vortex in a bilayer can be evaluated from the power balance $\eta v^2 B/\phi_0 =E^2/R_F$. Here  
the velocity $v$ of vortices with the areal density $B/\phi_0$ is related to the electric field $E$ in the core by the Faraday law $E=vB$, $R_F=R_\square B/B_{c2}$ is the flux flow sheet resistance, 
$R_\square=(d_1\sigma_1+d_2\sigma_2)^{-1}$ and 
$B_{c2}=\phi_0/2\pi\xi^2$. Hence $\eta$ acquires the conventional Bardeen-Stephen form $\eta=\phi_0^2/2\pi\xi^2R_\square$.  Expressing here $R_\square$ and $\xi$ in terms of the 
bilayer parameters yields
\begin{equation}
\eta=8\hbar(d_1N_1+d_2N_2)(T_{c0}-T).
\label{eto}
\end{equation}  
Here $\eta$ is independent of the mean free path in both N and S layers, consistent with the behavior of the Bardeen-Stephen drag coefficient $\eta_{BS}=8\hbar N(T_{c0}-T)$ per unit vortex length 
in the dirty limit at $T\approx T_c$~ \cite{kopnin}. 

\section{Current distribution in a partial vortex.}
\label{ap3}

\begin{figure}[tb]
	\includegraphics[width=9cm]{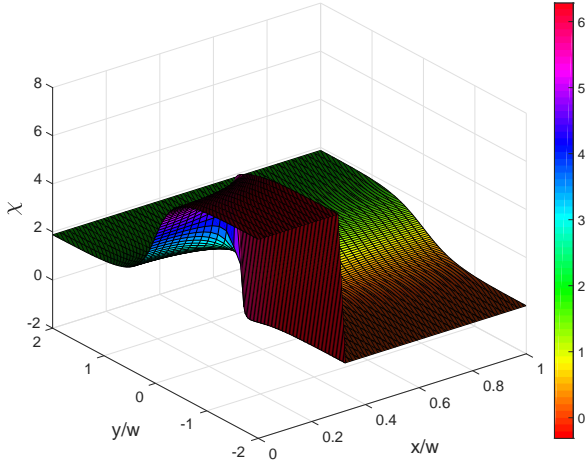}
   	\caption{ The surface plot of $\chi(x,y)$ calculated from Eq. (\ref{tha}) with the branch cut at $x=0$ and $-\infty <y<0$ for a vortex at $u=0.3w$.  
   }\label{fig9}
\end{figure}

A vortex in a thin film strip in which the London screening is negligible $(d\ll\lambda_L)$ can be described by the complex potential \cite{film,ahm}
\begin{gather}
g(z)=\chi(x,y) +ih(x,y)=i\ln\frac{\sin\frac{\pi}{2w}(z+u)}{\sin\frac{\pi}{2w}(z-u)},
\label{cp} \\
J_x-iJ_y=-\frac{\phi_0}{2\mu_0  \Lambda}\frac{dg}{dz},
\label{current}
\end{gather}
where $z=x+iy$, the strip is at $0<x<w$ and infinite along $y$, $\Lambda=\lambda_L^2/d$, and the vortex core is at $x=u,\, y=0$. 
Both components of the sheet current density $J_x(x,y)$ and $J_y(x,y)$ circulating around the vortex decrease exponentially 
over the length $w/\pi$ along the strip \cite{film}:  
\begin{gather}
J_x(x,y)=\frac{\phi_0}{4\mu_0\Lambda w}\biggl[\frac{\sinh \frac{\pi y}{w}}{\cosh\frac{\pi y}{w}-\cos\frac{\pi}{w}(x+u)} 
\nonumber \\
-\frac{\sinh \frac{\pi y}{w}}{\cosh\frac{\pi y}{w}-\cos\frac{\pi}{w}(x-u)}\biggr],
\label{jx} \\
J_y(x,y)=\frac{\phi_0}{4\mu_0\Lambda w}\biggl[\frac{\sin \frac{\pi }{w}(x-u)}{\cosh\frac{\pi y}{w}-\cos\frac{\pi}{w}(x-u)} 
\nonumber \\
-\frac{\sin \frac{\pi }{w}(x+u)}{\cosh\frac{\pi y}{w}-\cos\frac{\pi}{w}(x+u)}\biggr].
\label{jy}
\end{gather}
The function $\chi(x,y)$ in Eq. (\ref{cp}) is the phase of the order parameter which
is in turn the phase difference between the superconducting film and the overlayer with no current. 
The phase $\chi$ results in the Josephson energy: 
\begin{equation}
W_J=\frac{\hbar J_c}{2e}\int_0^w dx\int_{-L/2}^{L/2}dy[1-\cos\chi(x,y)].
\label{jose}
\end{equation}
Separation of the real part in Eq. (\ref{cp}) yields 
\begin{equation}
\chi=\tan^{-1}\frac{\tanh\frac{\pi y}{2w}}{\tan\frac{\pi}{2w}(x-u)}-\tan^{-1}\frac{\tanh\frac{\pi y}{2w}}{\tan\frac{\pi}{2w}(x+u)}.
\label{tha}
\end{equation}
As follows from Eq. (\ref{tha}), the vortex    
causes a nonzero phase $\chi_\infty (x)$ at $|y|\gg w/\pi$. If $y \to\infty$ Eq. (\ref{tha}) yields:
$\chi_\infty (x)=\pi u/w$ at $u<x<w$, $\chi_\infty (x)=\pi (u/w-1)$ at $0<x<u$,
and $\chi(x,\infty)=-\chi(x,-\infty)$. This form of $\chi_\infty(x)$ yields a discontinuity in the Josephson current density $\pm J_c\sin(\pi u/w)$ 
at $x=u$ across the bilayer. The discontinuity in $J_c\sin\chi$ can be removed by choosing a branch cut at $x=0$ and $-\infty <y<0$, giving
 \begin{gather}
\chi_\infty (x)=2\pi u/w, \qquad y\to\infty,
\label{th1} \\
\chi_\infty (x)=2\pi, \qquad 0<x<u, \quad y\to -\infty,
\label{th2} \\
\chi_\infty (x)=0, \qquad u<x<w, \quad y\to -\infty,
\label{th3}
\end{gather}
The function $\chi(x,y)$ is shown in Fig. \ref{fig9}. The constant phase difference $\chi_\infty$ at $y\gg w/\pi$ produces the 
Josephson energy proportional to the film area: 
\begin{equation}
W_J=\frac{\hbar J_c}{4e}Lw\left(1-\cos\frac{2\pi u}{w}\right)
\label{wjj}
\end{equation}

The finite phase difference $\chi_\infty$ at $y\gg w$ causes a transverse Josephson current $\sim J_cw$ which spreads through the layers 1 and 2. A self-consistent 
calculation of the phase distributions $\chi_1(\textbf{r})$ and $\chi_2(\textbf{r})$ in both layers requires solving the sine-Gordon equation obtained in Appendix \ref{ap5} along with Eq. (\ref{cp}).  

\section{Vortex inductive drag caused by a metallic overlayer}
\label{ap4}

A moving Pearl vortex produces the azimuthal vector potential $A_\varphi(x,y)$ outside a thin film \cite{pearl}:
\begin{equation}
A_{\varphi}(R)=\frac{\phi_{0}}{2\pi}\int_{0}^{\infty}\frac{J_{1}(kR)e^{-k|z|}}{1+2k\Lambda}dk,
\label{AP}
\end{equation}
where $\Lambda=\lambda_{L}^{2}/d_{1}$, $R=\sqrt{(x-u)^{2}+y^{2}}$, $u(t)$ is a time-dependent coordinate of the vortex core, and $J_1(x)$ is the Bessel function. 
The main contribution to the inductive drag comes from the region of radius $R\sim(d_{2}+d_{i})\ll \min(w,\Lambda)$ around the vortex, so the integral (\ref{AP}) is 
dominated by $k\Lambda\gg1$. In this case,
\begin{gather}
A_{\varphi}(R)=\frac{\phi_{0}}{4\pi\Lambda}\int_{0}^{\infty}\frac{J_{1}(kR)}{k}e^{-k|z|}dk=
\nonumber \\
\frac{\phi_0R}{4\pi\Lambda (|z|+\sqrt{R^{2}+z^{2}})}.
\label{Ap}
\end{gather}
The inductive electric field $E_\varphi(R)=-\dot{A}_\varphi$ produced by the moving vortex outside the film is then:
\begin{equation}
E_{\varphi}(x,y,z,t)=\frac{\dot{u}\phi_{0}(x-u)|z|}{4\pi \Lambda R\sqrt{R^{2}+z^{2}}(|z|+\sqrt{R^{2}+z^{2}})}.
\label{Ei}
\end{equation}

Let the vortex move with a slowly-varying velocity $\dot{u}(t)$ which only has low-frequency Fourier harmonics for 
which the skin depth,  $(\mu_0\sigma_2\omega)^{-1/2}$ is much larger than $d_2$. Then screening of a transverse electromagnetic field is negligible, so the inductive electric field $E_\varphi (R,z,t)$ of the vortex penetrates freely into the N overlayer. In this case Eq. (\ref{Ei}) can be used to calculate the ohmic power $P=\sigma_2\int_{V_2} E^2dxdydz$ in the N overlayer spaced by $d_{i}$ from the S film. Consider first the power density $p(z)=\sigma_2\int E^2 dxdy$ at the distance $z$ from the film and calculate the integral in polar coordinates centered in the moving vortex core:
\begin{equation}
\!\! p(z)=\frac{\dot{u}^{2}\phi_{0}^{2}\sigma_{2}z^{2}}{16\pi^{2}\Lambda^{2}}\int_{0}^{2\pi}\!\!\int_{0}^{\infty}\!\!\!\frac{R\cos^{2}\varphi d\varphi dR}{(R^{2}+z^{2})(|z|+\sqrt{R^{2}+z^{2}})^{2}}.
\label{pz}
\end{equation}
The main contribution to this integral comes from $R\sim z \sim d_i+d_2$, so the lateral size of the overlayer does not affect $p(z)$ if $d_i+d_2\ll w$. Integration in Eq. (\ref{pz}) gives:   
\begin{equation}
p(z)=\frac{\dot{u}^{2}\phi_{0}^{2}\sigma_{2}}{32\pi \Lambda^{2}}(\ln4-1).
\label{pzz}
\end{equation}
Since $p(z)$ turns out to be independent of $z$, the total power $P=\int_{d_i}^{d_i+d_2}\! p(z)dz$ is proportional to $d_2$:
\begin{equation}
P=\frac{\dot{u}^{2}\phi_{0}^{2}\sigma_{2}d_{2}}{32\pi \Lambda^{2}}(\ln4-1).
\label{Pi}
\end{equation}
Here $P$ is independent of the gap width $d_{i}$ as long as $ d_{i}+d_{2}\ll \min (w,\Lambda)$. In turn, the power $P$ can also be expressed in terms of the 
inductive vortex drag coefficient $\eta_2$ according to $\eta_2\dot{u}^2=P$. Hence,
\begin{equation}
\eta_{2}=\frac{\phi_{0}^{2}\sigma_{2}d_{2}d_{1}^{2}}{32\pi \lambda_{L}^{4}}(\ln4-1).
\end{equation}

\section{Current flow in a phase-unlocked bilayer.}
\label{ap5}

Consider a bilayer which carries the net current $I$:
\begin{equation}
I=d_{1}J_{1}+d_{2}J_{2}.
\label{Ia}
\end{equation}
Here the current densities in the layers 1 and 2 are related to the respective phase gradients as follows:
\begin{equation}
\textbf{J}_{1}=g_{1}\nabla\chi_{1},\qquad \textbf{J}_{2}=g_{2}\nabla\chi_{2}.
\label{curr} 
\end{equation}
If $J_1(x)$ varies slowly along the layer 1 over the length $\sim d_1$, the condition of current continuity becomes 
\begin{equation}
d_1\nabla\cdot{\bf J}_1+J_\perp=0, 
\label{jcont}
\end{equation}
where $J_\perp=J_{c}\sin\chi+R_{i}^{-1}V+C_{i}\partial_{t}V$ is the current density flowing through the interface.  Hence,
\begin{equation}
d_1g_1\nabla^2\chi_1=J_{c}\sin\chi+R_{i}^{-1}V+C_{i}\partial_{t}V,
\label{contin}
\end{equation}
where $V=(\hbar/2e)\partial_{t}\chi$ is the Josephson voltage, and $\chi=\chi_2-\chi_1$ is the phase difference between the layers. 
From Eqs (\ref{Ia}) and (\ref{curr}), it follows that $(d_{1}g_{1}+d_{2}g_{2})\nabla^{2}\chi_{1}=d_{2}g_{2}\nabla^{2}\chi$.
Substituting this into Eq. (\ref{contin}) yields the sine-Gordon equation for $\chi(\textbf{r},t)$: 
\begin{equation}
\omega_{J}^{-2}\partial_{tt}\chi+\tau\partial_{t}\chi=L_{J}^{2}\nabla^{2}\chi-\sin\chi,
\label{sge} 
\end{equation}
where 
$L_{J}^{2}=d_1d_2g_1g_2/(d_1g_1+d_2g_2)J_c$, $\omega_{J}^{2}=4e^{2}J_{c}/\hbar^{2}C_i$, $\tau=\hbar/2eR_i$, and 
$R_i$ and $C_i$ are the resistance and capacitance per unit area of the interface, respectively. 

A stationary solution of Eq. (\ref{sge}) that describes the current $I$ injected in the layer 1 at $x=0$ is:
\begin{equation}
\tan\frac{\chi}{4}=Ae^{-x/L_J},
\label{sol1}
\end{equation}
where $A$ is obtained from the boundary condition, $J_2(0)=0$, $I=-d_1g_1\chi'(0)$. Then Eq. (\ref{sol1}) yields, 
\begin{equation}
\frac{I}{4g_1d_1}(1+A^2)=\frac{A}{L_J}.
\label{bcj}
\end{equation}
The solution of Eq. (\ref{bcj}) for which $A=0$ at $I=0$ is:
\begin{equation}
A=\frac{2g_1d_1}{IL_J}-\sqrt{\left(\frac{2d_1g_1}{I L_J}\right)^2-1}
\label{A}
\end{equation}
Equations (\ref{sol1}) and (\ref{A}) yield Eq. (\ref{chi})


\begin{thebibliography}{99}

\bibitem{rev1}
S. Z. Butler {\it et al.} ASC Nano \textbf{7}, 2898 (2013).

\bibitem{rev2}
I. Bozovic and C. Ahn, Nat. Phys. \textbf{10}, 892 (2014). 

\bibitem{rev3}
Y. Saito, T. Nojima, and Y. Iwasa, Supercond. Sci. Technol. \textbf{29}, 093001 (2016).

\bibitem{rev4}
D. Huang and J. E. Hoffman, Annu. Rev. Cond. Mat. Phys. \textbf{8},  311 (2017). %hts on SrTi03

\bibitem{rev5}
T. Uchihashi, Supercond. Sci. Technol. \textbf{30}, 013003 (2017).

\bibitem{rev6}
D.-H. Lee, Annu. Rev. Cond. Mat. Phys. \textbf{9}, 261 (2018).

\bibitem{fese1}
Q. Y. Wang, L. Zhi, Z. Wen-Hao, Z. Zuo-Cheng, Z. Jin-Song, L. Wei, D. Hao, O. Yun-Bo, D. Peng, and C. Kai, Chin. Phys. Lett. \textbf{29}, 037402 (2012).

\bibitem{fese2}
 S. He {\it et al.}
 Nat. Mater. \textbf{12}, 605 (2013). %Tc=65K

\bibitem{fese3}
Y. Sun, W. Zhang, Y. Xing, F. Li, Y. Zhao, Z. Xia, L. Wang, X. Ma, Q. K. Xue, and J. Wang,  Sci. Rep. \textbf{4}, 06040 (2014). % Tc = 85K

\bibitem{fese4}
J.-F. Ge, Z.-L. Liu, C. Liu, C.-L. Gao, D. Qian, Q.-K. Xue, Y. Liu,
and J.-F. Jia, Nat. Mater. \textbf{14}, 285 (2015). %fese Tc = 109K

\bibitem{fese5}
S. Tan, Y. Zhang, M. Xia, Z. Ye, F. Chen, X. Xie, R. Peng, D. Xu, Q. Fan, H. Xu, J. Jiang, T. Zhang, X. Lai, T. Xiang, J. Hu, B. Xie, and D. Feng,  Nat. Mater. \textbf{12}, 634 (2013). % Interface-induced superconductivity and strain-dependent spin density waves in FeSe/SrTiO3 thin films

\bibitem{fese6}
R. Peng,  H. C. Xu, S. Y. Tan, H. Y. Cao, M. Xia, X. P. Shen, Z. C. Huang, C. H. P. Wen, Q. Song, T. Zhang, B. P. Xie, X. G. Gong, and D. L. Feng, Nat. Commun. \textbf{5}, 5044 (2014). %fese interface engineering

\bibitem{fese7}
J. J. Lee, F. T. Schmitt, R. G. Moore, S. Johnston, Y.-T. Cui, W. Li, M. Yi, Z. K. Liu, M. Hashimoto, Y. Zhang, D. H. Lu, T. P. Devereaux, D.-H. Lee, and Z.-X. Shen, Nature \textbf{515}, 245 (2014). %surface mode coupling

\bibitem{pb1}
C. Brun {\it et al.} Nat. Phys. \textbf{10}, 444 (2014).

\bibitem{pb2}
S. Yoshizawa, H. Kim, T. Kawakami, Y. Nagai, T. Nakayama, X. Hu, Y. Hasegawa, and T. Uchihashi, \prl~ \textbf{113}, 247004 (2014).

\bibitem{pb3}
D. Roditchev {\it et al.} Nat. Phys. \textbf{11}, 332 (2015).

\bibitem{tas2}
E. Navarro-Moratalla {\it et al.} Nat. Commun. \textbf{7}, 11043 (2015). 

\bibitem{kt1}
P.  Minnhagen, \rmp~\textbf{59}, 1001 (1987).

\bibitem{kt2}
J. M. Kosterlitz, Rep. Prog. Phys. \textbf{79}, 026001 (2016).

\bibitem{pfluct}
E. W. Carson, V. J. Emery, S. A. Kivelson, and D. Orgad.  {\it In Physics of Conventional and Unconventional
Superconductivity}, Vol. 2, ed. K.H. Bennemann and J.D. Ketterson, (Berlin, Heidelberg, New
York: Springer-Verlag), pp. 275-452, 2004).

\bibitem{agrev}
A. Gurevich, 
Annu. Rev. Cond. Mat. Phys, {\bf 5}, 35 (2014).

\bibitem{comp1}
B. Berg, D. Orgad, and S. A. Kivelson, \prb~\textbf{78}, 094509 (2008).

\bibitem{comp2}
G. Wachtel, A. Bar-Yaacov, and D. Orgad, \prb~\textbf{86}, 134531 (2012).

\bibitem{cooper}
L. N. Cooper, \prl~\textbf{6}, 689  (1961).

\bibitem{degennes}
G. Deutscher and P. G. de Gennes, {\it in Superconductivity}, Vol. 2, ed. R.D. Parks (New York: Marcel Dekker, Inc.) pp. 1005-1034 (1969).  

\bibitem{golub}
A. A. Golubov, M. Yu. Kupriyanov, and E. Il'ichev, \rmp~\textbf{76}, 411 (2004). 

\bibitem{belzig}
W. Belzig, C. Bruder, and G. Sch{\"o}n, \prb~\textbf{53}, 5727 (1996).

\bibitem{fominov} 
Ya. V. Fominov and M. V. Feigelman, \prb~\textbf{63}, 094518 (2001).

\bibitem{gol}
G. Brammertza, A. A. Golubov, P. Verhoeve, R. den Hartog, A. Peacock, and H. Rogalla, \apl ~\textbf{80}, 2955 (2002).

\bibitem{bi1}
I. L. Landau, D. L. Shapovalov, and I. A. Parshin, JETP Lett. \textbf{53}, 353 (1991).

\bibitem{bi2}
D. L. Shapovalov, JETP Lett. \textbf{60}, 199 (1994).

\bibitem{ga}
I. L. Landau and I. A. Parshin, Physica B \textbf{194-196}, 2339 (1994).

\bibitem{pb}
O. Bourgeois, A. Frydman, and R. C. Dynes, \prl~\textbf{88}, 186403 (2002); \prb~\textbf{68}, 092509 (2003).

\bibitem{nbn}
T. Shiino, S. Shiba, N. Sakai, T. Yamakura, L. Jiang, Y. Uzawa, H. Maezawa, and
S. Yamamoto, Supercond. Sci. Technol. \textbf{23}, 045004 (2010). 

\bibitem{koren}
O. Yuli, I. Asulin, O. Millo, D. Orgad, L. Iomin, and G. Koren, \prl~\textbf{101}, 057005 (2008). %LASCO bilayers

\bibitem{jjatunee}
A. J. Rimberg, T. R. Ho, C. Kurdak, J. Clarke, K. L. Campman, and A. C. Gossard, \prl~\textbf{78}, 2632 (1997).

\bibitem{kap}
N. Mason and A. Kapitulnik, \prb~\textbf{65}, 220505(R) (2002).

\bibitem{jjatunet}
K.-H. Wagenblast, A. van Otterlo, G. Sch{\"o}n, and G. T. Zim{\' a}nyi, \prl~\textbf{79}, 2730 (1997).

\bibitem{finkels}
K. Michaeli and A. M. Finkel'stein, \prl~\textbf{97}, 117004 (2006); \prb~\textbf{76}, 064506 (2007).

\bibitem{magtune}
V. G. Kogan, \prb~\textbf{75}, 064514 (2007). 

\bibitem{is1}
A. Gozar and I. Bozovic, Physica C \textbf{521-522}, 38 (2016).

\bibitem{is2}
X. Y. Tee, T. Ito, T. Ushiyama, Y. Tomioka, I. Martin, and C. Panagopoulos,
\prb~ \textbf{95}, 054516 (2017).

\bibitem{coulomb1}
A. M. Finkelstein, Physica B~ \textbf{197}, 636 (1994).

\bibitem{coulomb2}
Y. Oreg, P. W. Brouwer, B. D. Simons, and A. Altland, \prl~\textbf{82}, 1269 (1999).

\bibitem{pearl}
J. Pearl, \apl~ {\bf 5}, 65 (1964); A. L. Fetter and P. C. Hohenberg, Phys. Rev. \textbf{159}, 330 (1967).

\bibitem{ren1}
J. Tobochnik and G. V. Chester, \prb~\textbf{20}, 3761 (1979).

\bibitem{ren2}
H. Weber and P. Minnhagen, \prb~\textbf{37}, 5986 (1988).

\bibitem{renexp}
A. T. Fiory, A. F. Hebard, and W. I. Glaberson, \prb~\textbf{28}, 5075 (1983).

\bibitem{bktm}
E. J. K{\"o}nig, A. Levchenko, I. V. Protopopov, I. V. Gornyi, I. S. Burmistrov, and A. D. Mirlin, \prb~\textbf{92}, 214503 (2015).

\bibitem{amplf}
A. Erez and Y. Meir, \prb~\textbf{88}, 184510 (2013).

\bibitem{kopnin}
N. B. Kopnin, {\it Theory of Nonequilibrium Superconductivity.} (Oxford Univ. Press, New York, 2001).

\bibitem{hu}
C.-R. Hu, \prb~\textbf{6}, 1756 (1972).

\bibitem{beasley}
M. R. Beasley, J. E. Mooij, and T. P. Orlando, \prl~\textbf{42}, 1165 (1979).

\bibitem{disordbkt}
J. Um, B. J. Kim, P. Minnhagen, M. Y. Choi, and S-I. Lee, \prb~\textbf{74}, 094516 (2006).   % strong reduction of Tc by disortder Tc = (0.22-0.265)J

\bibitem{broad}
L. Benfatto, C. Castellani, and T. Giamarchi, \prb~\textbf{80}, 214506 (2009).

\bibitem{tun}
J. Zasadzinski. {\it Tunneling spectroscopy of conventional and unconventional superconductors. In The Physics of Superconductors}, 
(Ed. K.H. Bennemann and J.B. Ketterson, Springer, Berlin, Heidelberg, New York)  v. 1, p. 591 (2003).

\bibitem{blstm1}
Z. Long, M. D. Stewart, T. Kouh, and J. M. Valles, \prl~\textbf{93}, 257001 (2004);

\bibitem{blstm2}
Z. Long, M. D. Stewart, and J. M. Valles, \prb~\textbf{73}, 140507(R) (2006).

\bibitem{stmPb}
L. Serrier-Garcia, J. C. Cuevas, T. Cren, C. Brun, V. Cherkez, F. Debontridder,
D. Fokin, F. S. Bergeret, and D. Roditchev, \prl ~\textbf{110}, 157003 (2013).

\bibitem{dynes1}
R. C. Dynes, V.  Narayanamurti, and J. P. Garno, \prl ~{\bf 41}, 1509 (1978). 

\bibitem{dynes2}
R. C. Dynes, J. P.  Garno, J. P.  Hertel, and T. P.  Orlando, \prl ~{\bf 53}, 2437 (1984).

\bibitem{inelast}
T. P. Devereaux and D. Belitz, \prb ~{\bf 44}, 4587 (1991).

\bibitem{coulomb}
D. A. Browne, K. Levin, and K. A. Muttalib, \prl ~{\bf 58}, 156 (1987).

\bibitem{anis}
A. N. Bennett, Phys. Rev. \textbf{140}, A1902 (1965).

\bibitem{larkin}
A. I. Larkin and Yu. N. Ovchinnikov, Zh. Exp. Teor. Fiz. {\bf 61}, 2147 (1971) [ JETP {\bf 34}, 1144 (1972) ].

\bibitem{balatski}
A. V. Balatskii, I. Vekhter, and J-X. Zhu, \rmp ~{\bf 78}, 373 (2006).

\bibitem{meyer}
J. S. Meyer and B. D. Simons, \prb ~{\bf 64}, 134516 (2001).

\bibitem{arnold}
E. L. Wolf and G. B. Arnold, Phys. Rep. \textbf{91}, 31 (1982).

\bibitem{feigel}
M. A. Skvortsov and M. V. Feigel'man, Zh. Exp. Teo. Fiz. \textbf{144}, 560 (2013) [JETP \textbf{117}, 487 (2013)].

\bibitem{kubo}
A. Gurevich and T. Kubo, \prb \textbf{96}, 184515 (2017).

\bibitem{jja1} 
R. S. Newrock, C. J. Lobb, U. Geigenm{\" u}ller, and M. Octavio, Solid State Physics, \textbf{54}, 263 (2000).

\bibitem{jja2}
R. Fazio and H. van der Zant, Phys. Rep. \textbf{355}, 235 (2001).

\bibitem{KKL}
K. K. Likharev, {\it Dynamics of Josephson Junctions and Circuits} (Gordon and Breach, New York, 1986).

\bibitem{granul}
S. John and T. C. Lubensky, \prb~\textbf{34}, 4815 (1986)

\bibitem{granrev}
I. S. Beloborodov, A. V. Lopatin, V. M. Vinokur, and K. B. Efetov, \rmp~\textbf{79}, 469 (2007).

\bibitem{fluct}
A. Larkin and A. Varlamov, {\it Theory of Fluctuations in Superconductors} (Oxford University Press, New York, Hong Kong, Madrid, Toronto, 2009). 

\bibitem{lam1}
S. D. Adrian, M. E. Reeves, S. A. Wolf, and V. Z. Kresin, \prb~\textbf{51}, 6800 (1995).

\bibitem{lam2}
X. Leyronas and R. Combescot, \prb~\textbf{54}, 3482 (1996).

\bibitem{lam3}
E. J. Nicol and J. P. Carbotte, \prb~\textbf{71}, 054501 (2005).

\bibitem{comm}
The renormalized vortex energy  should, in principle, be proportional to $\zeta_1 d_1\sigma_2+\zeta_2 d_2\sigma_2$ with two different constants $\zeta_1$ and $\zeta_2$. 
However, we use here the simpler form given by Eq. (\ref{bkt}), assuming that the effects of weak localization in the highly conductive layer 2 would be much less pronounced than  
in the dirtier layer 1. There are other uncertainties due to, for example, surface and interface scattering which can be essential in ultra-thin films.   

\bibitem{film}
G. Stejic, A. Gurevich, E. Kadyrov, D. Christen, R. Joynt, and D. C. Larbalestier, \prb~\textbf{49}, 1274 (1994).

\bibitem{ahm}
 A. Sheikhzada and A. Gurevich, \prb~ \textbf{95}, 214507 (2017).

\bibitem{gv}
A. Gurevich and V. M. Vinokur, \prl~\textbf{100}, 227007 (2008).

\bibitem{embon}
L. Embon, Y. Anahory, A. Suhov, D. Halbertal, J. Cuppens, A. Yakovenko, A. Uri, Y. Myasoedov, M. L. Rappaport, M. E. Huber, A. Gurevich, and E. Zeldov,  Sci. Rep. \textbf{5}, 7598 (2015).

\bibitem{fv1}
L. F. Chibotaru and V. H. Dao, \prb~\textbf{81}, 020502(R) (2010).

\bibitem{fv2}
Y. Tanaka,  H. Yamamori, T. Yanagisawa, T. Nishio, and S. Arisawa, Physica C \textbf{548}, 44 (2018).

\bibitem{mlphase}
A. Gurevich and V. M. Vinokur, \prl~\textbf{97}, 137003 (2006).
 
\bibitem{stanf}
J. M. Graybeal and M. R. Beasley, \prb~\textbf{29},  4167 (1984). 
 
\bibitem{int1}
R. P. Robertazzi, A. W. Kleinsasser, R. B. Laibowitz, R. H. Koch, and K. G. Stawiasz, \prb~\textbf{46}, 8456 (1992).

\bibitem{int2}
J. W. Ekin, S. E. Russek, C. C. Clickner, and B. Jeanneret, \apl~\textbf{62}, 369 (1993). 
 
\bibitem{Ash}
N. W. Ashcroft and N. D. Mermin, {\it Solid State Physics} (Brooks/Cole, Belmont, 1976).

\bibitem{clem}
J. R. Clem, J. Low Temp. Phys. \textbf{18}, 427 (1975).


\end{thebibliography}
\end{document}